\documentclass[twocolumn]{aastex62}
\usepackage{empheq}

\shorttitle{Properties of the KISS GPs}
\shortauthors{Brunker et al.}

\def\etal{et al.~}

\begin{document}

\title{PROPERTIES OF THE KISS GREEN PEA GALAXIES}

\correspondingauthor{Samantha Brunker}
\email{sbrunker@indiana.edu}

\author[0000-0001-6776-2550]{Samantha W. Brunker}
\affiliation{Department of Astronomy, Indiana University, 727 East Third Street, Bloomington, IN 47405, USA}
 
\author[0000-0001-8483-603X]{John J. Salzer}
\affiliation{Department of Astronomy, Indiana University, 727 East Third Street, Bloomington, IN 47405, USA}

\author[0000-0001-9165-8905]{Steven Janowiecki}
\affiliation{University of Texas at Austin, McDonald Observatory, TX 79734, USA}

\author[0000-0001-8518-4862]{Rose A. Finn}
\affiliation{Department of Physics, Siena College, 515 Loudon Road, Loudonville, NY 12211, USA}

\author[0000-0003-3367-3415]{George Helou}
\affiliation{Infrared Processing and Analysis Center, California Institute of Technology, Pasadena, CA 91125, USA}

\begin{abstract}
Green Peas are a class of extreme star-forming galaxies at intermediate redshifts, originally discovered via color-selection using multi-filter, wide-field survey imaging data \citep{Cardamone2009}.  They are commonly thought of as being analogs of high-redshift Ly$\alpha$-emitting galaxies.  The defining characteristic of Green Pea galaxies is a high-excitation nebular spectrum with very large equivalent width lines, leading to the recognition that Green Pea-like galaxies can also be identified in samples of emission-line galaxies.  Here we compare the properties a sample of [O~III]-selected star-forming galaxies (z = 0.29-0.41) from the KPNO International Spectroscopic Survey (KISS) with the color-selected Green Peas.  We find that the KISS [O~III]-selected galaxies overlap with the parameter space defined by the color-selected Green Peas; the two samples appear to be drawn from the same population of objects.   We compare the KISS Green Peas with the full H$\alpha$-selected KISS star-forming galaxy sample (z $<$ 0.1) and find that they are extreme systems.  Many appear to be young systems at their observed look-back times (3-4 Gyr), with more than 90\% of their rest-frame B-band luminosity coming from the starburst population.   We compute the volume density of the KISSR Green Peas at z = 0.29-0.41 and find that they are extremely rare objects.   We don't see galaxies as extreme as the KISSR Green Peas in the local Universe, although we recognize several lower-luminosity systems at z $<$ 0.1.  

\end{abstract}

\keywords{Emission-Line Galaxies; Abundances; Star Formation; Starburst; Evolution}

\section{Introduction} \label{sec:intro}

One of the foremost topics in modern extragalactic research is the ongoing attempt to understand the key physical processes that occurred during the era of reionization.  During this period of time in the early Universe, the intergalactic medium (IGM) went from being neutral and opaque for photons capable of ionizing Hydrogen to ionized and transparent.  Reionization appears to have been complete by z $\sim$ 6 \citep[e.g.,][]{fan2006,mcgreer2015}, though determining exactly when it began is much less constrained, and depends critically on the nature of the ionizing sources.  Proposed candidates for the source of the ionizing radiation include star-forming galaxies \citep[SFG; e.g.,][]{robertson2010}, active galactic nuclei \citep[AGN; e.g.,][]{haiman1998,madau2015}, and quasars \citep[e.g.,][]{madau2004}.  

The most commonly accepted narrative is that star-forming galaxies are the main contributors to reionization. This idea is complicated by the fact that star-forming regions are typically surrounded by large HI column densities which prevent the ionizing radiation produced by hot stars from escaping into the IGM. Low and intermediate mass galaxies may get around this problem if they have a fully-ionized interstellar medium, or one perforated by optically thin tunnels by which the ionizing radiation could escape \citep{jaskot2013,nakajima2014,rivera-thorsen2015,izotov2018b}.  The production and escape of ionizing radiation in star-forming galaxies is not yet fully understood, although major advances in the past 5 -- 10 years suggest that we are on the path toward a more complete understanding.  

As it is difficult to observe high redshift galaxies that leak Lyman-continuum radiation (LyC, $\lambda_{rest}$ $<$ 912 \AA), studies of lower-redshift LyC emitters are necessary in order to understand how ionizing radiation escapes from SFGs.  A class of compact SFGs known as Green Peas \citep[GPs;][]{Cardamone2009} have become popular targets for these observations because of their apparent similarities to high-z SFGs \citep[e.g., low metallicities and high specific star formation rates;][]{nakajima2014,Izotov2011,henry2015}.  A significant fraction of Green Peas show high Ly$\alpha$ escape fractions from 1-50\% \citep[e.g.,][]{henry2015,yang2017,verhamme2017,jaskot2017,jaskot2019,izotov2020}.  The Green Peas also include some of the only known LyC leaking SFGs in the local Universe \citep[e.g.,][]{izotov2016a,izotov2016b,izotov2017,izotov2018a}, and all of the systems with $f_{esc}$(LyC) $>$ 5\%.   This makes Green Peas important systems for studying the escape mechanisms of ionizing radiation. 

 The Green Peas were originally discovered by citizen scientists as part of the Galaxy Zoo galaxy classification project.  Following this discovery, \citet{Cardamone2009} published the first sample of GPs which were color selected from the Sloan Digital Sky Survey \citep[SDSS;][]{sdss}.  However, this is not the only way to discover such objects.  A key property of GPs is their extremely strong [O~III]$\lambda$5007 emission lines.  However, the SDSS color-based selection will only allow for the detection of GP-like objects in a restricted redshift range (see \S 3). More traditional emission-line selection methods (e.g., objective-prism/grism surveys, narrow-band surveys) can allow for the discovery of GP-like objects over much broader redshift ranges \citep{hoyos2005, kakazu2007}.  For example, a sample of GP-like galaxies was discovered via their [O~III] emission lines as part of the KPNO International Spectroscopic Survey \citep[KISS;][]{salzer2000, Salzer2009}.  These [O~III]-selected star-forming galaxies appear to have properties similar to the \citet{Cardamone2009} sample.   

In the current study we focus on three issues.   First, are the KISS [O~III]-selected emission-line galaxies (ELGs) truly analogous to the color-selected GP galaxies?  We attempt to answer this question by directly comparing the properties of the KISS objects with the \citet{Cardamone2009} GP sample.  Second, we explore the nature and evolutionary status of the GP-like galaxies.  Finally, we ask whether there are any GP galaxies in the very local (z $<$ 0.1) Universe, and attempt to address the question of what the GPs seen at intermediate redshifts look like today.   To address these latter two items, we utilize the full KISS catalog to create a local comparison sample of actively star-forming galaxies and to look for nearby GP analogs.

In this paper we present the properties of the KISS [O~III]-detected star-forming galaxies.  The sample selection and new observational data for the galaxies are presented in Section~\ref{sec:kissGP}. The comparison of the KISS [O~III]-detected sample with the \citet{Cardamone2009} sample is presented in Section~\ref{sec:KISSasGP}.  In Section~\ref{sec:EvolutionaryStatus} we compare the KISS [O~III]-detected sample with the low-redshift KISS H$\alpha$-detected sample as well as discuss the evolutionary status of the [O~III]-detected galaxies. An analysis of the volume densities of the KISS [O~III]-detected galaxies is presented in Section~\ref{sec:VolumeDensities}, and a discussion of Green Peas in the local Universe is presented in Section~\ref{sec:GPsNow}.  Our findings are summarized in Section~\ref{sec:conclusions}.
 
All derived distance-dependent quantities assume a standard cosmology of H$_o$ = 70 km/s/Mpc, $\Omega_\Lambda$ = 0.73 and $\Omega_M$ = 0.27 throughout the paper.

\section{The [O~III]-selected KISS Galaxies} \label{sec:kissGP}
\subsection{The KISS Red Emission-Line Galaxies}

The GP-like galaxies being studied in the current paper were all discovered in the KPNO International Spectroscopic Survey \citep[KISS;][]{salzer2000}.  KISS employed a low-dispersion objective prism on the 0.61 m Burrell Schmidt\footnote{The Burrell Schmidt telescope of the Warner and Swasey Observatory is operated by Case Western Reserve University} telescope on Kitt Peak to carry out a comprehensive survey of ELGs in the nearby Universe.   The objective-prism spectra covered two distinct wavelength ranges: $\lambda\lambda$ 6400-7200 \AA\ (KISS red, selected primarily by the H$\alpha$ line \citep{salzer2001, gronwall2004b, jangren2005a}) and $\lambda\lambda$ 4800-5500 \AA\ (KISS blue, selected primarily by the [O~III]$\lambda$5007 line; \citep{salzer2002}).  All of the GP-like galaxies were detected in the KISS red portion of the survey.

A program of ``quick-look" follow-up spectroscopy was carried out by members of the KISS collaboration.  These spectra were used for determining accurate redshifts and for ascertaining the activity type of each galaxy (e.g., star forming {\it versus} AGN).   Spectroscopic observations for the KISS ELGs are presented in \citet{wegner2003, gronwall2004a, jangren2005b, salzer2005b}; additional unpublished spectral data also exist.  See \citet{hirschauer2018} for a recent summary of the full spectroscopic follow-up of KISS.   All 2157 KISS red ELGs from the first two survey catalogs \citep[][hereafter KR1 and KR2, respectively]{salzer2001, gronwall2004b} possess follow-up spectra.
The current study focuses primarily on the galaxies in the KR1 and KR2 lists.

\begin{figure*}[ht]
\centering
\includegraphics[width=0.48\linewidth]{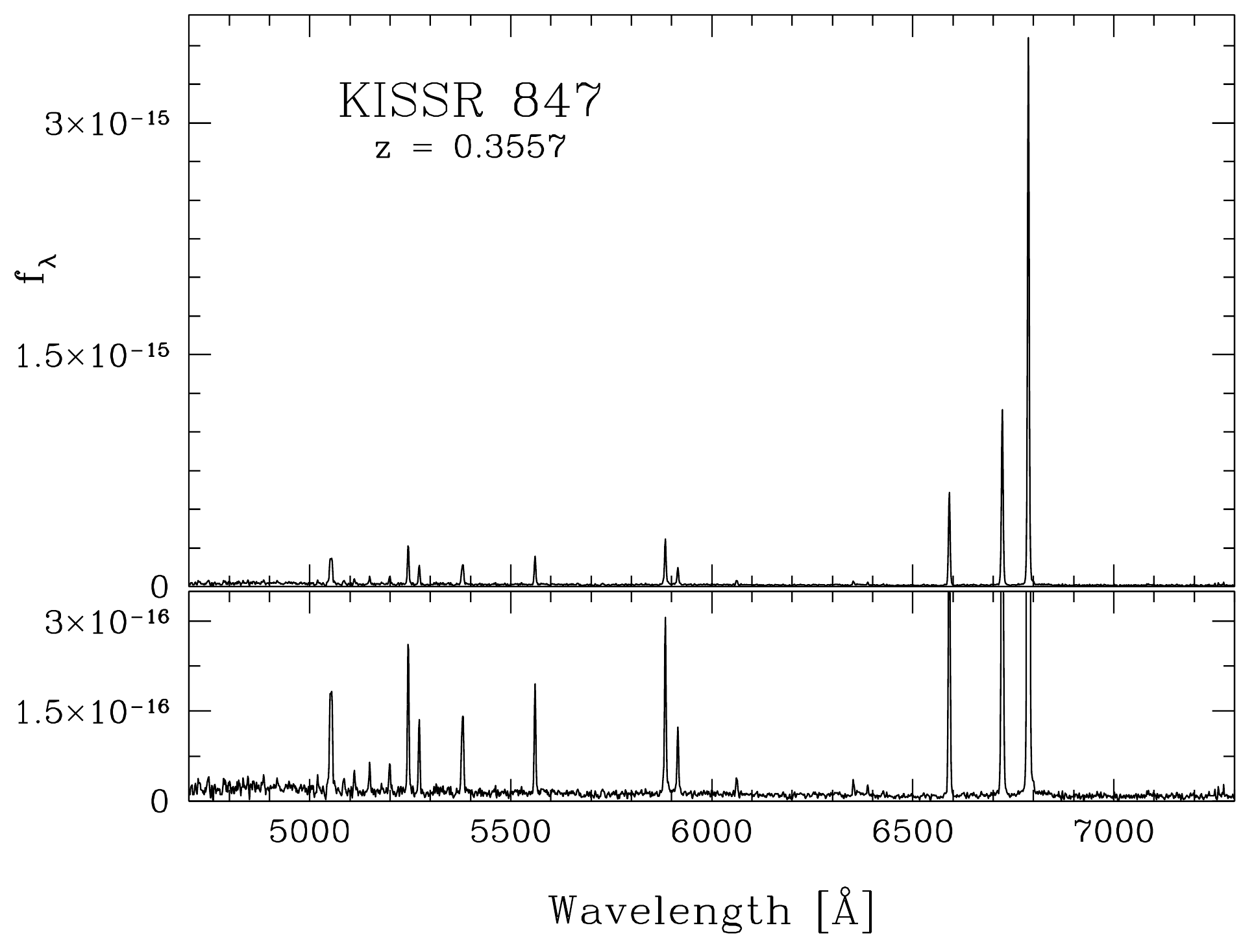}\quad\includegraphics[width=0.48\linewidth]{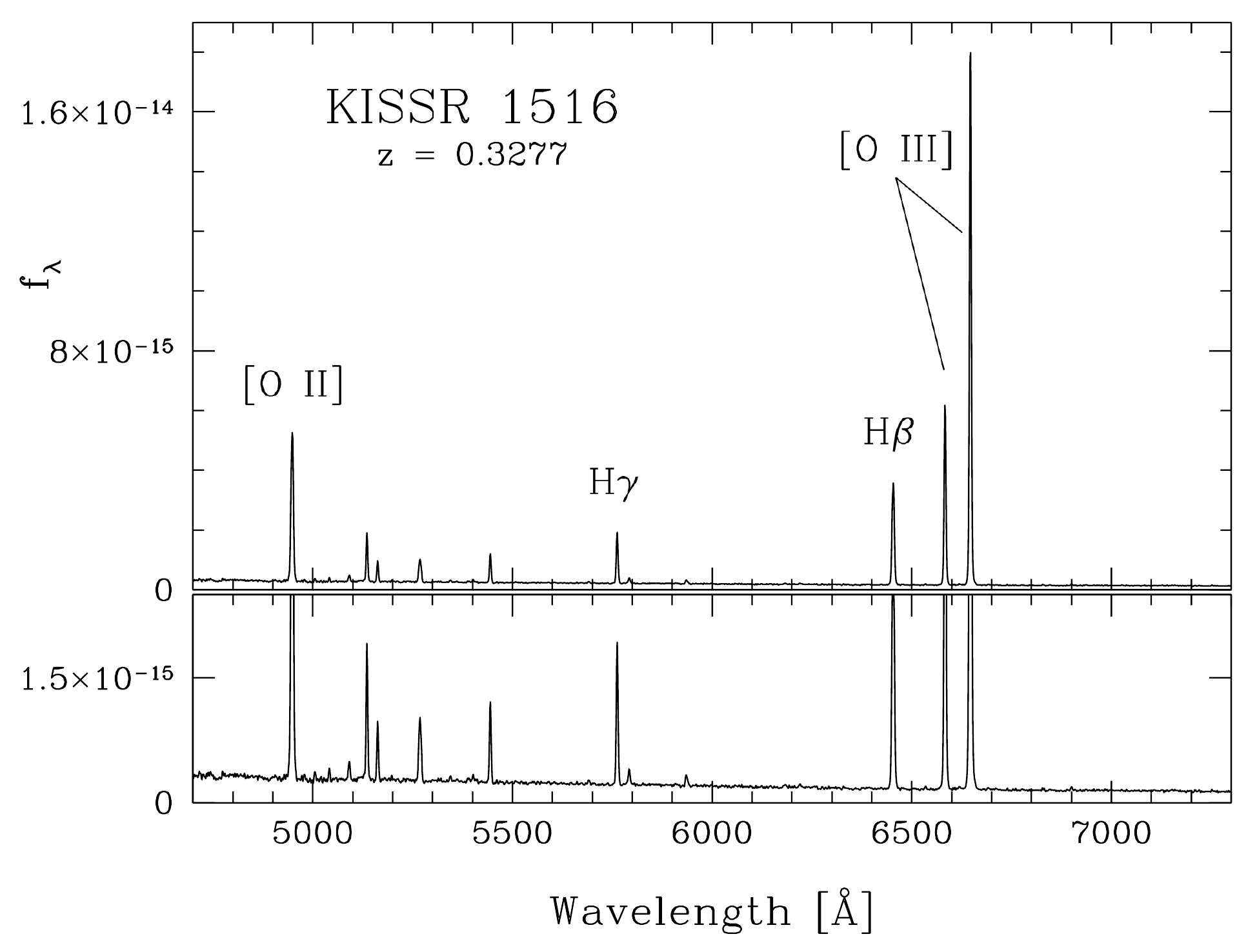}
\\[\baselineskip]
\includegraphics[width=0.48\linewidth]{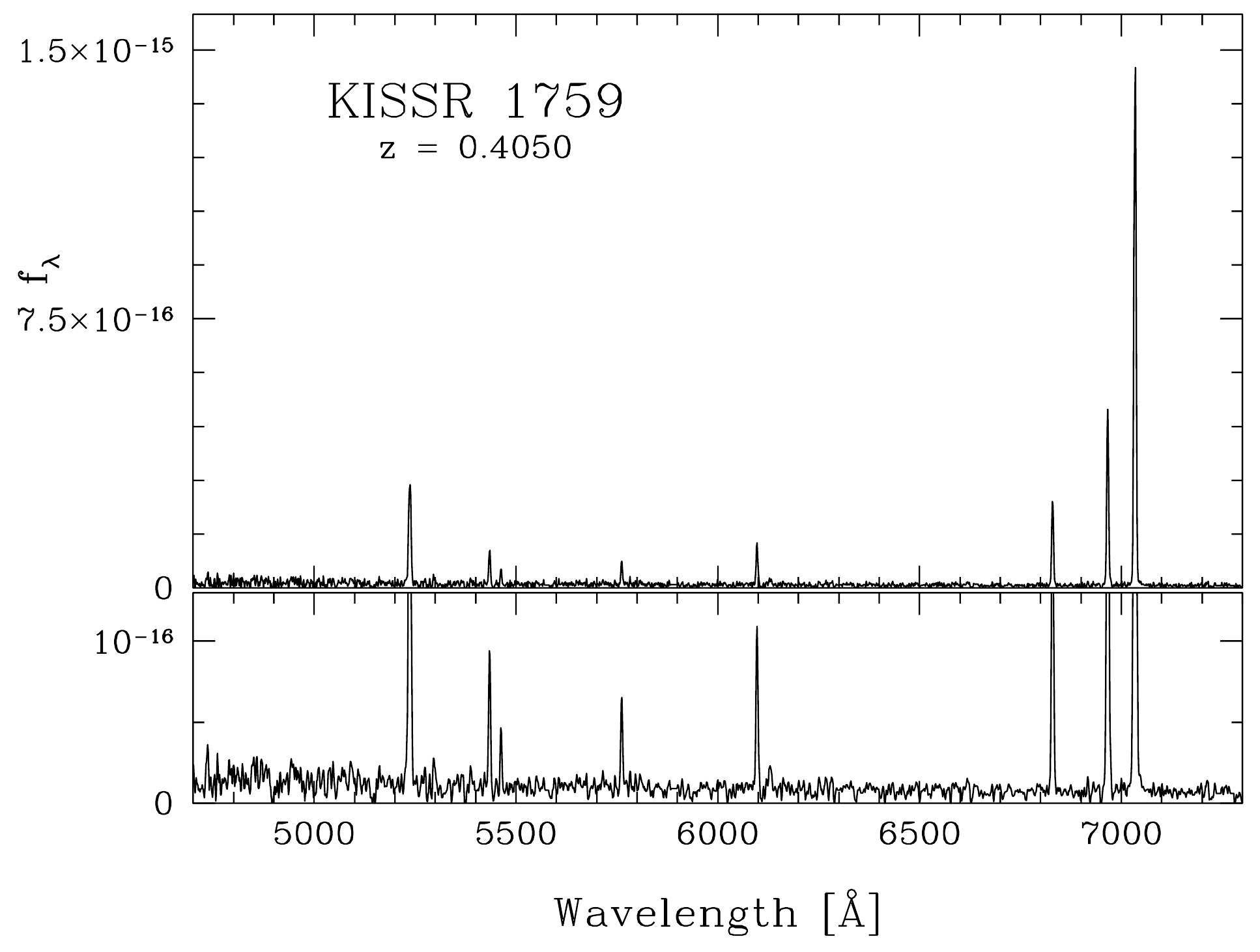}\quad\includegraphics[width=0.48\linewidth]{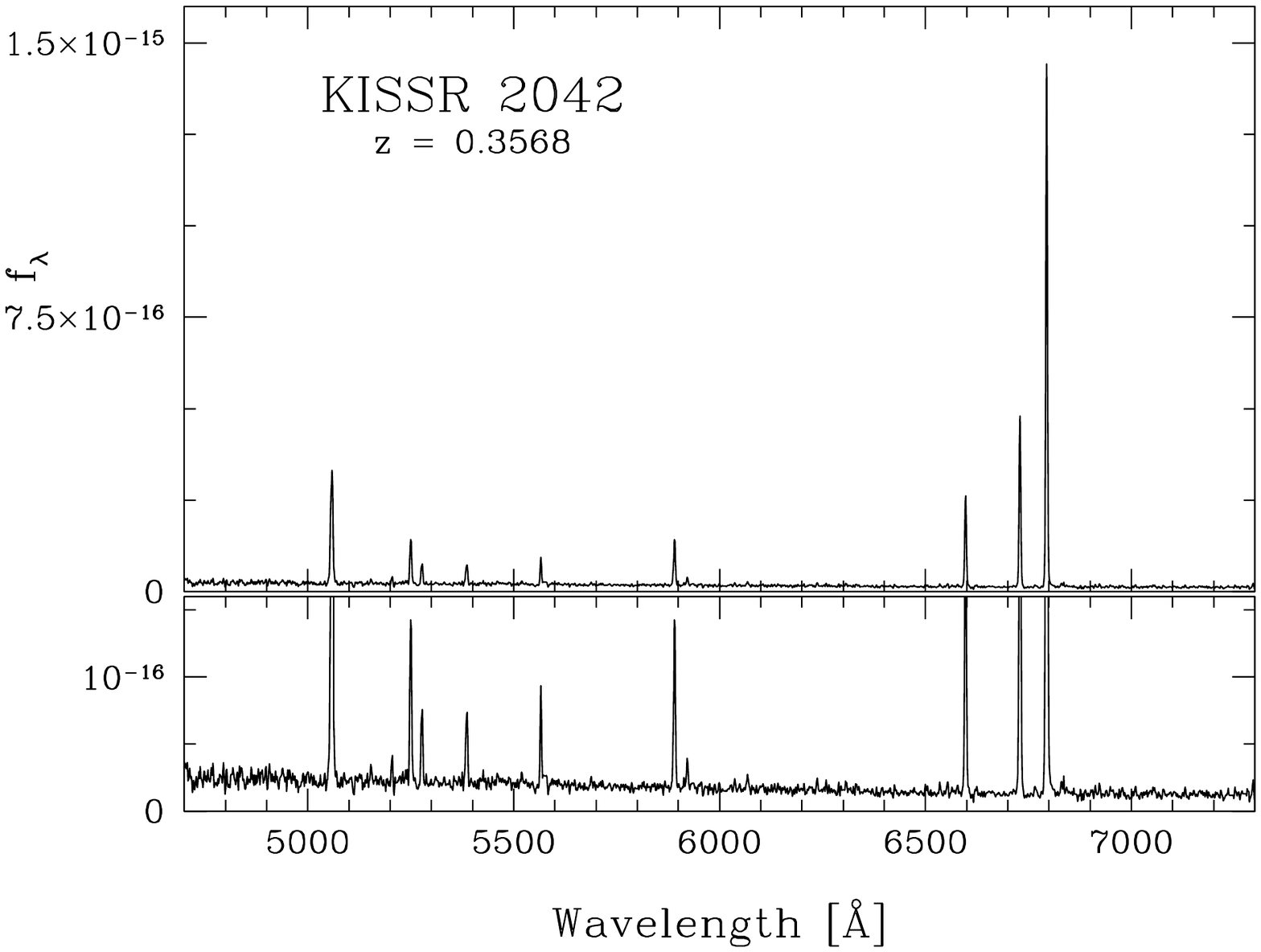}
\caption{\footnotesize Spectra of four KISSR GP-like galaxies.  Key emission lines are labeled in the upper right spectrum.  All exhibit the strong [O~III]$\lambda\lambda$5007,4959 lines characteristic of the Green Pea galaxies.  KISSR 847 (upper left) is extreme in all aspects, exhibiting very high EW [O~III] lines but very weak [O~II]$\lambda\lambda$3729,3726.  The lower sub-panel of each spectrum shows a clipped version of the y-axis in order to allow the reader to see the weaker emission lines, including [O~III]$\lambda$4363 just redward of H$\gamma$.  We also possess spectra that reach redward to the location of the H$\alpha$ and [N~II] lines (not shown) that allow us to confirm the star-forming nature of these galaxies. 
}
\label{fig:spec}
\end{figure*}

The vast majority of the KISS red (hereafter KISSR) galaxies were detected via the H$\alpha$ emission line, resulting in a sample of ELGs with redshifts between 0.0 and 0.095 (limited by the survey filter).  However, roughly 2\% were detected by strong [O~III]$\lambda$5007 emission that was redshifted into the KISSR filter bandpass.  These objects, which have redshifts between 0.29 and 0.42, tend to have large equivalent width emission lines and high excitation (large [O~III]/H$\beta$) values.  

In a preliminary analysis, \citet{Salzer2009} found that 15 of the 38 [O~III]-detected KISSR ELGs were star-forming systems based on their locations in line diagnostic diagrams \citep[e.g.,][]{bpt, vo1987}; the remaining [O~III]-detections were all AGNs, mainly Seyfert 2 galaxies.  Two of the star-forming KISSR galaxies (KISSR 169 and 980) do not have characteristics that mimic the GP galaxies and are not discussed further in this study.

Most of the [O~III]-detected star-forming galaxies posssessed high-excitation spectra that were reminiscent of dwarf star-forming galaxies despite their high luminosities (M$_B$ $\sim$ $-$20 to $-$22).  Using a strong-line metallicity indicator  \citep[e.g.,][]{melbourne2002, salzer2005a}, \citet{Salzer2009} showed that the [O~III]-detected KISSR ELGs were located more than a factor of ten below local star-forming galaxies in a metallicity-luminosity diagram.  None of the spectra available at that time were of sufficient quality to allow for the derivation of direct abundances.

\begin{deluxetable*}{ccccccccc}
\tabletypesize{\footnotesize}
\tablewidth{0pt}
\tablecaption{New NIR and Spitzer Observations of the KISSR [O~III]-detected Galaxies \label{tab:irdata}}

\tablehead{
 \colhead{KISSR} &  \colhead{J} &  \colhead{H} &  \colhead{IRAC 3.6} &\colhead{IRAC 4.5} &  \colhead{IRAC 5.7} &  \colhead{IRAC 7.9} &  \colhead{MIPS 24} &  \colhead{MIPS 70} \\
 & [mag] & [mag] & \colhead{[mJy]} & \colhead{[mJy]} & \colhead{[mJy]} & \colhead{[mJy]} & \colhead{[mJy]} & \colhead{[mJy]} \\
 (1)  & (2)  & (3)  & (4)  & (5)  & (6)  & (7)  & (8)  & (9) 
}

\startdata
\ 225 & 18.31 $\pm$ 0.04 &  17.53 $\pm$ 0.04 & 75.4 $\pm$ 1.2 & 84.1 $\pm$ 1.6 & 78.0 $\pm$ 4.8 & 399.9 $\pm$ 10.4 & 1567 $\pm$ 33 & 11930 $\pm$ 345 \\
\ 560 & 19.96 $\pm$ 0.16 &  18.67 $\pm$ 0.09 & --- &  --- &  --- &  --- &  --- &  --- \\
\ 847 & ---              &  19.54 $\pm$ 0.23 & 22.2 $\pm$ 0.8 & 30.6 $\pm$ 1.3 & 43.2 $\pm$ 4.0 & 66.3 $\pm$ 8.7 & \ 567 $\pm$ 31 & --- \\
 1038 & 19.40 $\pm$ 0.06 &  18.52 $\pm$ 0.06 & 28.6 $\pm$ 0.9 & 28.7 $\pm$ 1.2 & 41.4 $\pm$ 4.1 & 86.5 $\pm$ 8.2 & 1762 $\pm$ 28 & 14060 $\pm$ 394 \\
 1290 & 18.76 $\pm$ 0.07 &  18.20 $\pm$ 0.06 & --- &  --- &  --- &  --- &  --- &  --- \\
 1508 & 18.77 $\pm$ 0.07 &  17.99 $\pm$ 0.05 & --- &  --- &  --- &  --- &  --- &  --- \\
 1516 & 18.60 $\pm$ 0.04 &  18.12 $\pm$ 0.05 & 54.1 $\pm$ 1.0 & 77.9 $\pm$ 1.5 & 96.3 $\pm$ 4.1 & 310.7 $\pm$ 9.0 & 4495 $\pm$ 30 & 24400 $\pm$ 406 \\
 1759 & 20.15 $\pm$ 0.14 &  19.46 $\pm$ 0.10 & 40.3 $\pm$ 0.9 & 91.6 $\pm$ 1.6 & 243.7 $\pm$ 4.7 & 576.5 $\pm$ 9.6 & 2919 $\pm$ 29 & 9273 $\pm$ 296 \\
 1791 & ---              &  17.73 $\pm$ 0.13 & --- &  --- &  --- &  --- &  --- &  --- \\
 1825 & 17.94 $\pm$ 0.04 &  16.96 $\pm$ 0.05 & --- &  --- &  --- &  --- &  --- &  --- \\
 1953 & 18.68 $\pm$ 0.06 &  17.90 $\pm$ 0.04 & 49.6 $\pm$ 1.0 & 55.7 $\pm$ 1.4 & 56.2 $\pm$ 4.1 & 149.1 $\pm$ 8.8 & 2084 $\pm$ 29 & 18120 $\pm$ 316 \\
 2005 & 18.69 $\pm$ 0.07 &  17.95 $\pm$ 0.08 & --- &  --- &  --- &  --- &  --- &  --- \\
 2042 & 19.95 $\pm$ 0.08 &  19.45 $\pm$ 0.09 & 11.8 $\pm$ 0.7 & 11.2 $\pm$ 1.1 & 23.2 $\pm$ 4.3 & --- & 302 $\pm$ 28 & ---
\enddata
\end{deluxetable*}

We illustrate the spectral characteristics of the KISSR [O~III]-detected star-forming galaxies in Figure~\ref{fig:spec}.  These spectra were obtained recently as part of a study to measure accurate metallicities of the KISS GP-like galaxies (Brunker \etal in preparation).  Compared to the KISS ``quick-look" spectra described above, these data possess substantially higher S/N and will allow, in many cases, the derivation of direct abundances.  The presence of [O~III]$\lambda$4363 temperature-sensitive auroral line is clearly visible in all of the examples shown.  It is worth stressing that the KISSR [O~III]-detected galaxies have spectra reminiscent of the the GP galaxies \citep[e.g.,][]{Cardamone2009, Izotov2011}.  

\subsection{New IR Observations and Derivation of Stellar Masses}

The mass of the stellar component of a galaxy is a key parameter for interpreting its nature.  Traditional methods for deriving stellar masses include (1) using optical luminosities and colors combined with modeled mass-to-light ratios or (2) spectral energy distribution (SED) fitting using optical broad-band magnitudes \citep{bell2001,bell2003,walcher2011}.  However, for systems with intense star formation, such as the ones in the current study, these standard mass-estimation methods can be highly inaccurate.  This is true for two reasons.  First, in extreme starbursts the optical broad-band luminosities can be enhanced by huge amounts due to the luminous young stars.  The population of newly-formed stars often contributes very little to the total stellar mass despite dominating the light output.  Second, nebular line emission can dramatically change the broad-band colors of these galaxies (case in point: the Green Peas).  Since the simple M/L prescriptions used in method (1) above are based on stellar emission alone, this can lead to large discrepancies.  In such cases, the best approach to deriving reliable stellar masses involves utilizing observations in the near-IR, where the contribution from the young stars does not dominate so severely. 

In order to provide better determinations of the stellar masses for the [O~III]-detected KISSR galaxies, we have undertaken new observations.  These new data include both ground-based near-IR (NIR) imaging as well as space-based mid-IR (MIR) and far-IR (FIR) observations with the Spitzer Space Telescope.

None of the star-forming [O~III]-detected KISSR galaxies are bright enough to have been detected by the 2-Micron All-Sky Survey \citep[2MASS;][]{skrutskie2006}.   Therefore, we carried out NIR imaging observations using WHIRC \citep[WIYN High-resolution InfraRed Camera;][]{whirc} on the WIYN\footnote{The WIYN Observatory is a joint facility of the University of Wisconsin-Madison, Indiana University, the University of Missouri, Purdue University, and the National Optical Astronomy Observatory} 3.5-m telescope.   The observational methods and data reduction procedures mirror those described in \citet{janowiecki2014}.  In total, eleven of the thirteen [O~III]-detected KISSR galaxies were observed in both the J (1.25 $\mu$m) and H (1.65 $\mu$m) filters, while the remaining two were observed only in H.  Our NIR magnitudes for the KISSR [O~III]-detected galaxies are listed in Table~\ref{tab:irdata}.

\begin{figure*}[ht]
\centering
\includegraphics[width=0.49\linewidth]{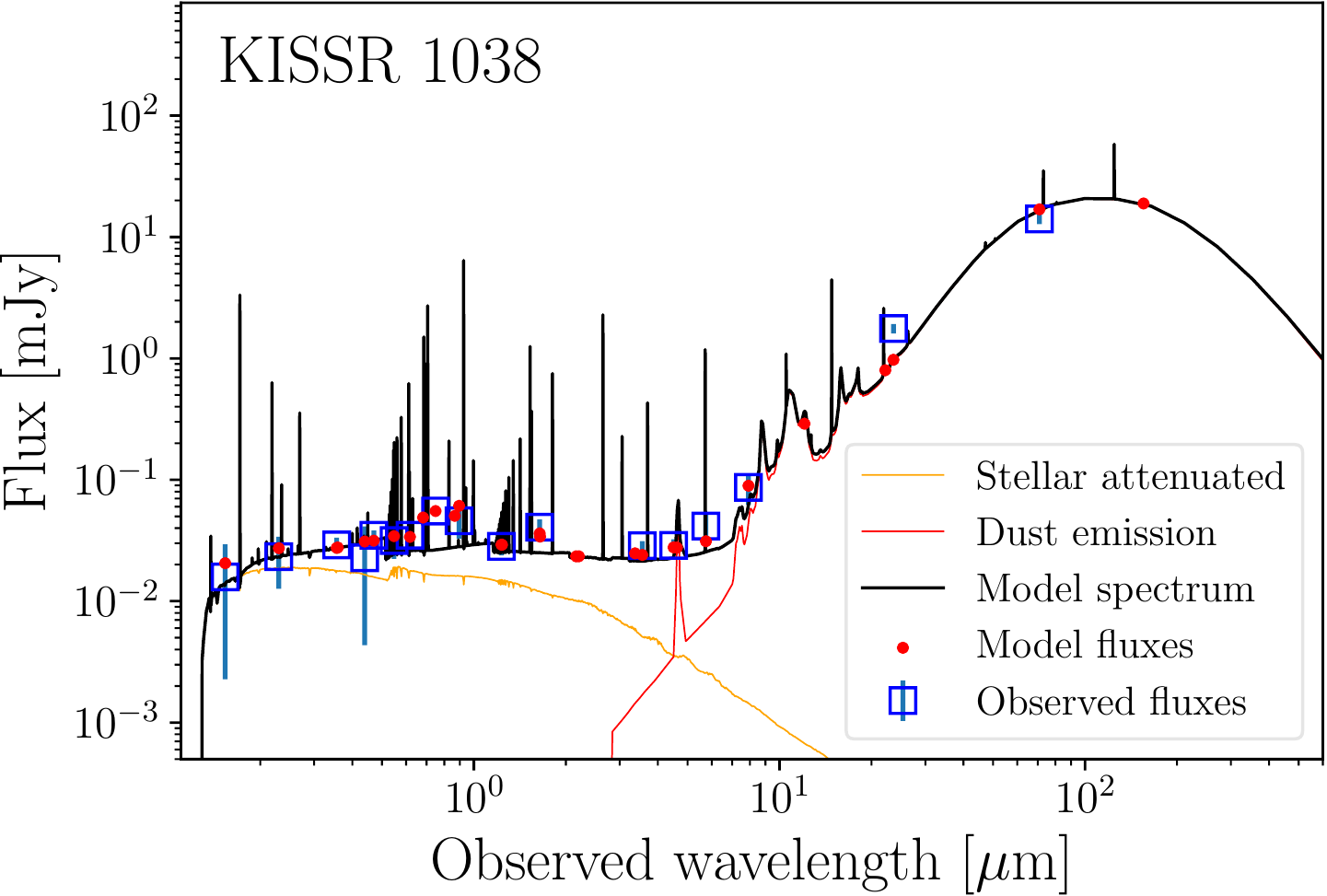}\quad\includegraphics[width=0.49\linewidth]{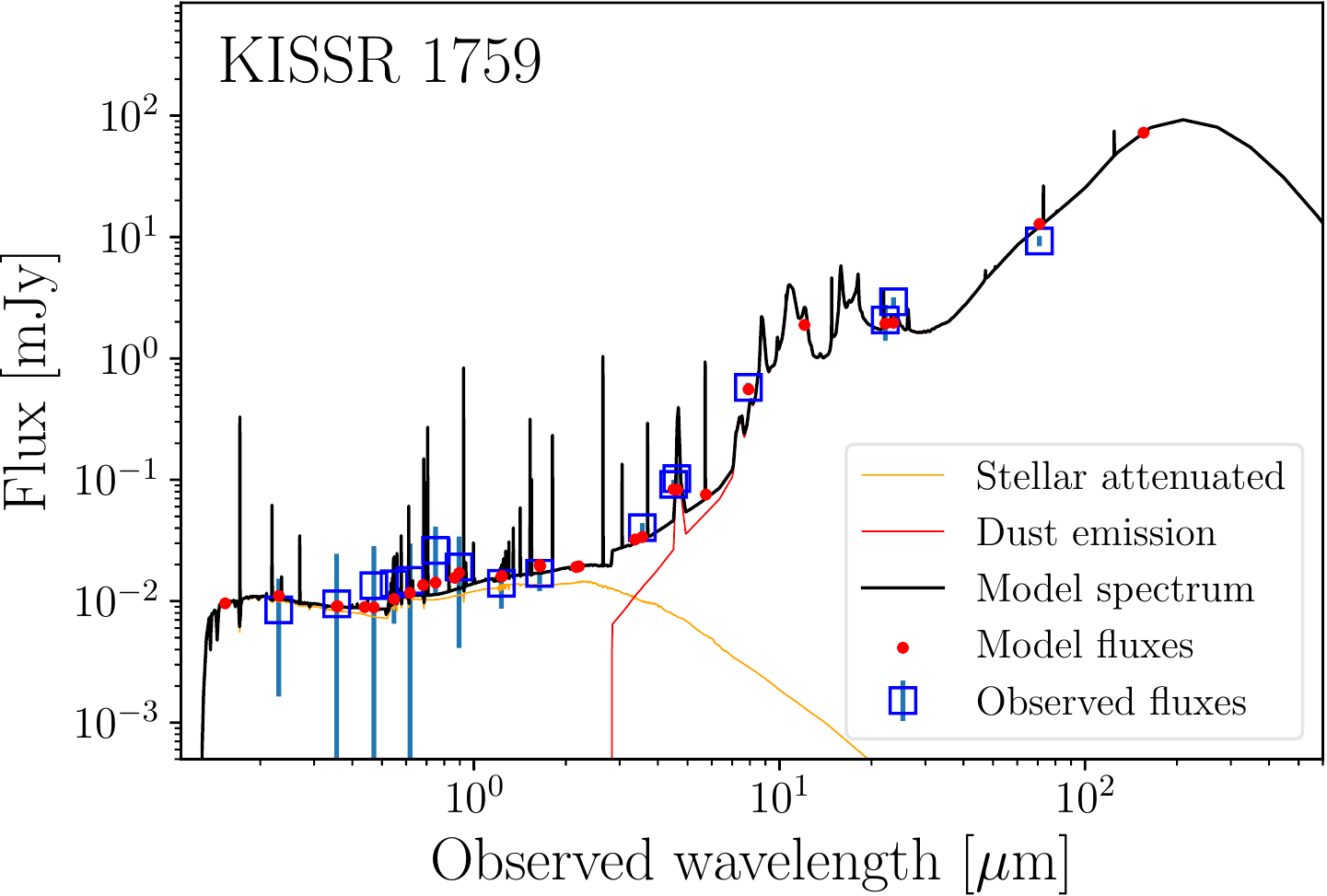}
\\[\baselineskip]
\includegraphics[width=0.49\linewidth]{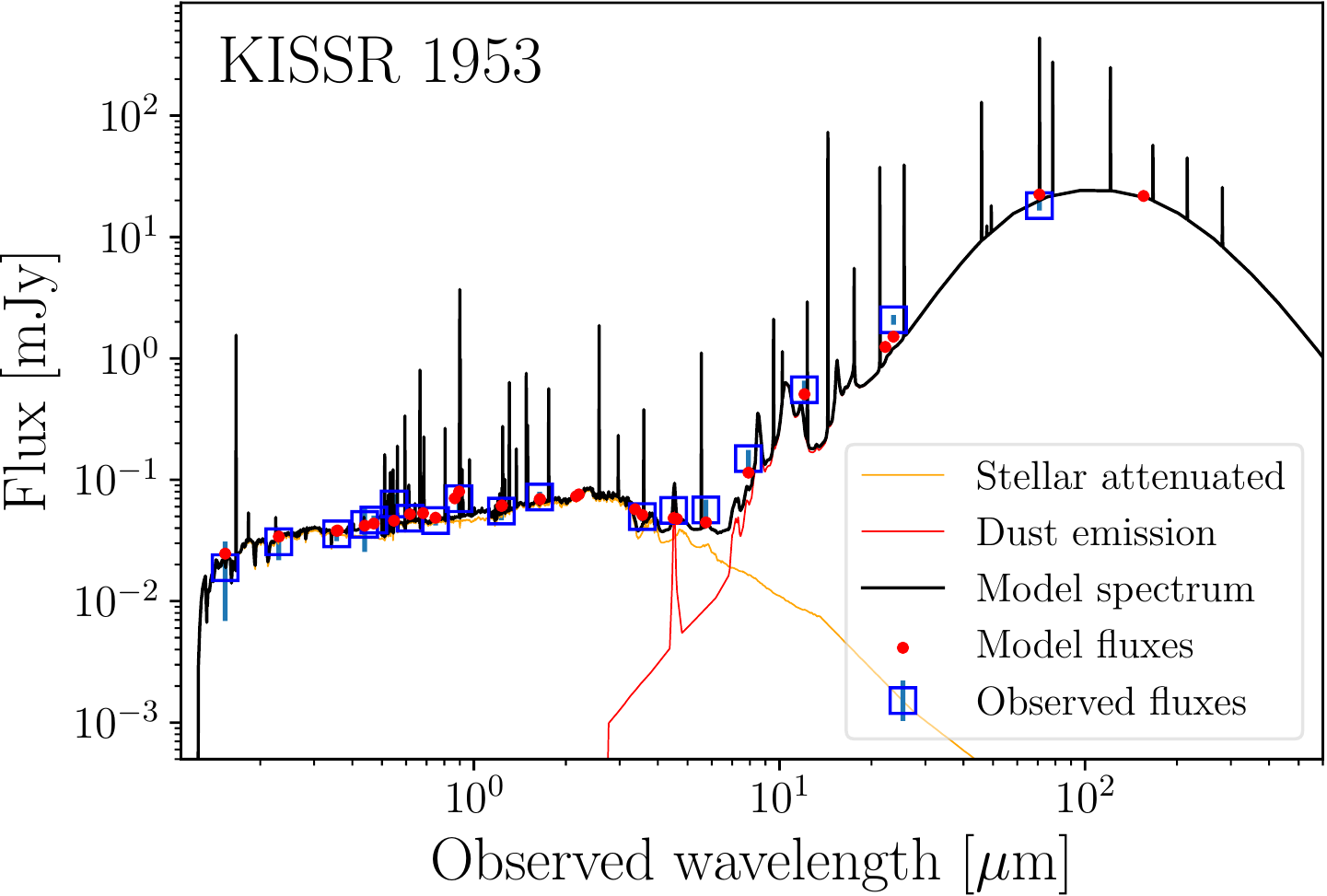}\quad\includegraphics[width=0.49\linewidth]{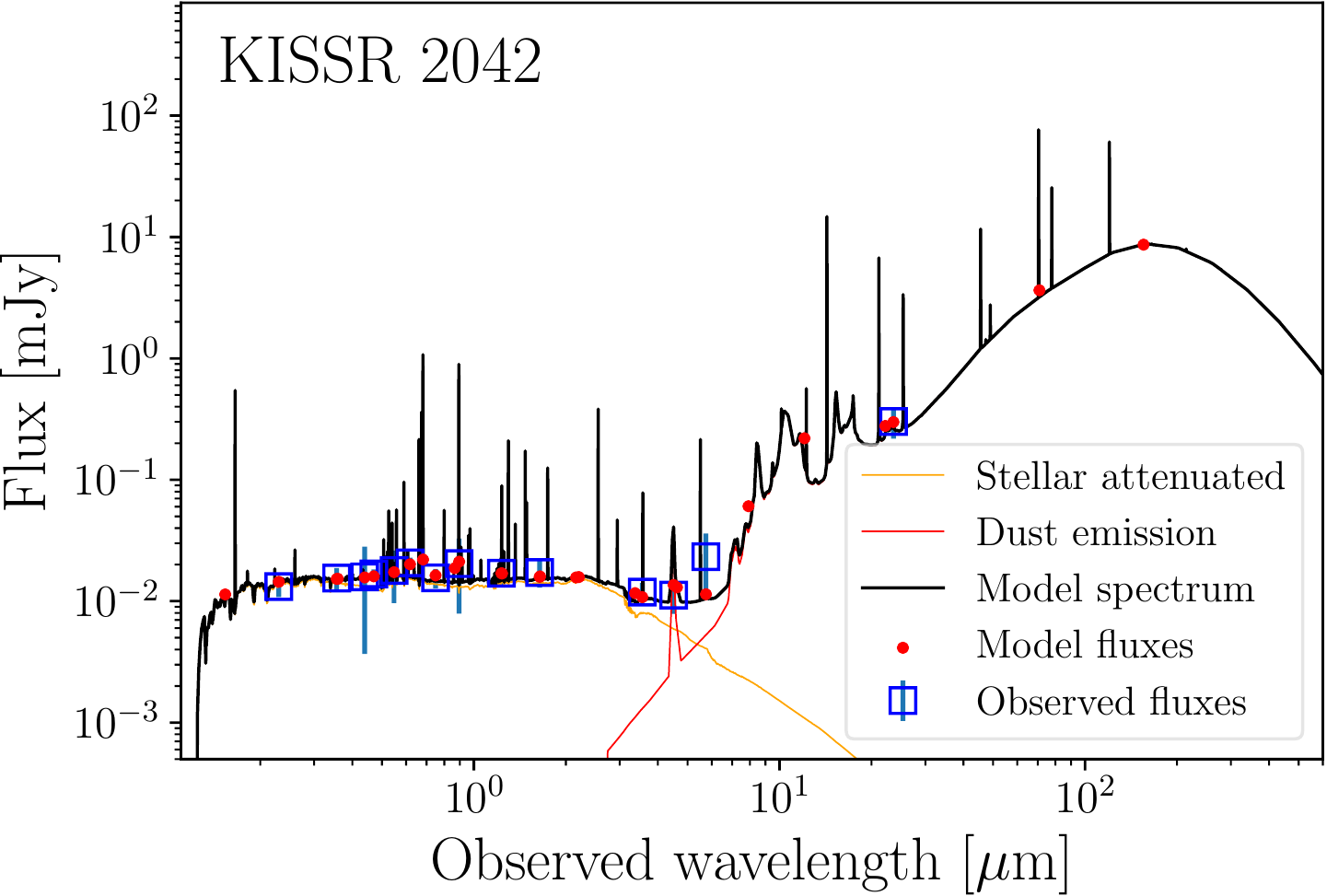}

\caption{\footnotesize Observed spectral energy distributions (SEDs) and model fits for four [O III]-detected KISSR galaxies.   The blue boxes represent the locations of measured flux points, which range from the UV (GALEX FUV and NUV fluxes) to the FIR (Spitzer MIPS fluxes).  The CIGALE model fits for the stellar component (orange curve), the dust component (red curve), and the composite model (stellar, dust, and nebular emission; black curve) are shown.  
}
\label{fig:sed}
\end{figure*}

Seven of the KISSR [O~III]-detected star-forming galaxies were observed with the Spitzer Space Telescope\footnote{This work is based in part on observations made with the Spitzer Space Telescope, which was operated by the Jet Propulsion Laboratory, California Institute of Technology under a contract with NASA.} using both the IRAC (3.6, 4.5, 5.7 and 7.9 $\mu$m) and MIPS (24 and 70 $\mu$m) instruments.  The observations were carried out in 2009 as part of a Director's Discretionary Time request; the objects selected for observation were chosen to cover the range of observed optical luminosities.  We used the MOsaicker and Point source EXtractor (MOPEX) software tool\footnote{https://irsa.ipac.caltech.edu/data/SPITZER/docs/data-analysistools/tools/mopex/} to reduce both the IRAC and MIPS data.  For the IRAC data, we followed the the {\it Spitzer Data Analysis Cookbook} for ``Point Source Photometry from IRAC Images''\footnote{https://irsa.ipac.caltech.edu/data/SPITZER/docs/data-analysistools/cookbook/12/}.  We applied an array-dependent photometric correction to the corrected Basic Calibration Data (CBCD) images because our sources are blue. We used the Astronomical Point Source EXtraction (APEX) software to perform point source fitting on individual images using APEX multiframe point source extraction mode, and we used a point response function (PRF) map provided by {\it Spitzer Space Center}.  We used the total flux from APEX.

\begin{deluxetable*}{ccccccccrcc}
\tabletypesize{\footnotesize}
\tablewidth{0pt}
\tablecaption{Properties of KISSR [O~III]-detected Galaxies \label{tab:prop}}

\tablehead{
 \colhead{KISSR} &  \colhead{RA} &  \colhead{DEC} &  \colhead{B$_o$} &  \colhead{(B$-$V)$_o$} & \colhead{z} & \colhead{EW$_{[O~III]}$} & \colhead{M$_B$} &  \colhead{log(M$_*$)} &  \colhead{log(O/H)+12} &  \colhead{log(SFR)}  \\
  &  \colhead{degrees} &  \colhead{degrees} &  &  &  & \AA &  &  \colhead{M$_\odot$}  &  &  \colhead{M$_\odot$/yr} \\
 (1)  & (2)  & (3)  & (4)  & (5)  & (6)  & (7)  & (8)  & (9)  & (10) & (11)
}
 
\startdata
 \ 225 &  197.54067 &   29.29527 &  19.81 &   0.41 &   0.35822 & \ 102.4 &  -21.61 &    9.73 &    8.58 &    1.61 \\
 \ 560 &  217.45028 &   29.48553 &  21.47 &   0.89 &   0.35787 & \ 405.7 &  -19.95 &    8.99 &    7.80 &    0.80 \\
 \ 847 &  235.98020 &   29.46394 &  21.89 &   0.64 &   0.35575 &  1896.0 &  -19.52 &    8.86 &    7.65 &    1.36 \\
  1038 &  247.14056 &   29.32112 &  20.57 &   0.46 &   0.41001 & \ 731.1 &  -21.20 &    9.69 &    8.20 &    1.63 \\
  1290 &  185.60078 &   43.18965 &  20.13 &   0.58 &   0.30496 & \ 272.7 &  -20.89 &    9.44 &    8.25 &    0.97 \\
  1508 &  198.67200 &   43.72399 &  19.90 &   0.49 &   0.29397 & \ 413.1 &  -21.02 &    9.55 &    8.30 &    1.01 \\
  1516 &  198.95600 &   43.57503 &  19.51 &   0.15 &   0.32766 & \ 685.6 &  -21.69 &    9.67 &    8.17 &    1.32 \\
  1759 &  215.15889 &   43.72554 &  21.79 &   0.77 &   0.40504 &  1129.0 &  -19.95 &    9.32 &    7.97 &    1.02 \\
  1791 &  217.42047 &   43.90243 &  20.43 &   0.06 &   0.35923 & \ 213.1 &  -21.00 &   10.15 &    8.41 &    1.37 \\
  1825 &  219.08469 &   43.88378 &  19.56 &   0.27 &   0.33112 & \ 155.5 &  -21.67 &   10.25 &    8.60 &    1.26 \\
  1953 &  231.59900 &   43.00455 &  19.92 &   0.54 &   0.36882 & \ 427.2 &  -21.58 &    9.62 &    8.20 &    1.42 \\
  2005 &  235.69358 &   43.89937 &  20.02 &   0.26 &   0.30805 & \ 345.1 &  -21.03 &    9.10 &    8.46 &    1.31 \\
  2042 &  237.44008 &   43.05719 &  21.00 &   0.24 &   0.35680 & \ 953.2 &  -20.42 &    8.98 &    7.77 &    1.14 
\enddata
\end{deluxetable*}

For the MIPS 24\micron \ imaging, we created the image mosaics following the reduction procedure outlined in \citet{finn2010} and the {\it Spitzer Data Analysis Cookbook} for ``MIPS 24\micron \ mosaic of a $z=0.7$ cluster''\footnote{https://irsa.ipac.caltech.edu/data/SPITZER/docs/data-analysistools/cookbook/28/}.  We flattened the basic calibration data (BCD) images and then created a mosaic with MOPEX.  We extracted source photometry using APEX in single-frame point source extraction mode, fitting the PRF in each image.  Again, we use the total flux value calculated by APEX.  The procedure for creating the mosaic and extracting fluxes was similar for the MIPS 70\micron\  images\footnote{https://irsa.ipac.caltech.edu/data/SPITZER/docs/data-analysistools/cookbook/31/}\footnote{https://irsa.ipac.caltech.edu/data/SPITZER/docs/data-analysistools/cookbook/34/}, except that we started with the filtered BCD (FBCD) images that are available from the $Spitzer$ archive.  The FBCD images correct for striping that arises from differences in detector response, an effect that is not removed in the BCD images. The FBCD images yield reliable photometry for point sources with fluxes below 0.5~Jy, and they improved our ability to detect the KISS galaxies at 70\micron.  The measured mid- and far-IR fluxes for the seven KISSR [O~III]-detected galaxies are listed in Table~\ref{tab:irdata}.

In addition to these new observations, the WISE all-sky survey \citep{wise} was sensitive enough to detect several of the KISSR [O~III]-detected star-forming galaxies in one or more MIR bands (3.4, 4.6, 12 and 22 $\mu$m).  All six of the KISSR [O~III]-detected galaxies not observed by Spitzer were detected with WISE in at least one band, meaning the we have important NIR and MIR flux points for all of the [O~III]-detected KISSR galaxy SEDs.  For three KISSR galaxies with both Spitzer and WISE data, the agreement between the measured fluxes is quite good.  

Detailed SED fits like those illustrated in Figure~\ref{fig:sed} were used to determine the stellar masses of these systems. We utilized the Code Investigating Galaxy Emission (CIGALE) software \citep{noll2009}. CIGALE generates a grid of SEDs based on various theoretical models and compares with observed SEDs to determine the best values for each parameter in the models. Our wavelength coverage from UV to MIR (or in some cases, FIR) allows CIGALE to self-consistently account for any dust absorption and re-emission. Modules were employed which describe the nebular emission and absorption \citep{inoue2011}, dust attenuation \citep{cardelli1989}, and thermal dust emission \citep{dale2014}, using stellar population synthesis models \citep{bruzual2003} and a \citet{salpeter1955} initial mass function. Our fitting process and determination of stellar mass is more fully described in Appendix B of \citet{hirschauer2018} and in Section 3 of \citet{janowiecki2017}.  We stress that the extreme star-forming nature of these types of systems makes the determination of accurate stellar masses extremely difficult.   The inclusion of the NIR and MIR data is essential for constraining the stellar masses.  We present our mass estimates in Table~\ref{tab:prop}.   Characteristic uncertainties in the mass determinations are 0.2-0.3 dex in log(M$_*$) (i.e., 30-50\% uncertainties in the mass).

Example broad-wavelength-coverage SEDs overplotted with the CIGALE model fits are illustrated in Figure~\ref{fig:sed}.  The data included in the SEDs ranges from the UV (GALEX fluxes) through to the FIR (Spitzer fluxes).  The model fits include separate spectral plots for the stellar component (orange) and the dust component (red).  The nebular emission component is not plotted separately, but is included in the composite model spectrum (black).  The blue boxes indicate the locations of measured fluxes for each galaxy.  The location of the boxes vary from galaxy to galaxy, depending on the specific data set available for each galaxy.

\begin{figure*}[ht]
\centering
\includegraphics[width=\linewidth]{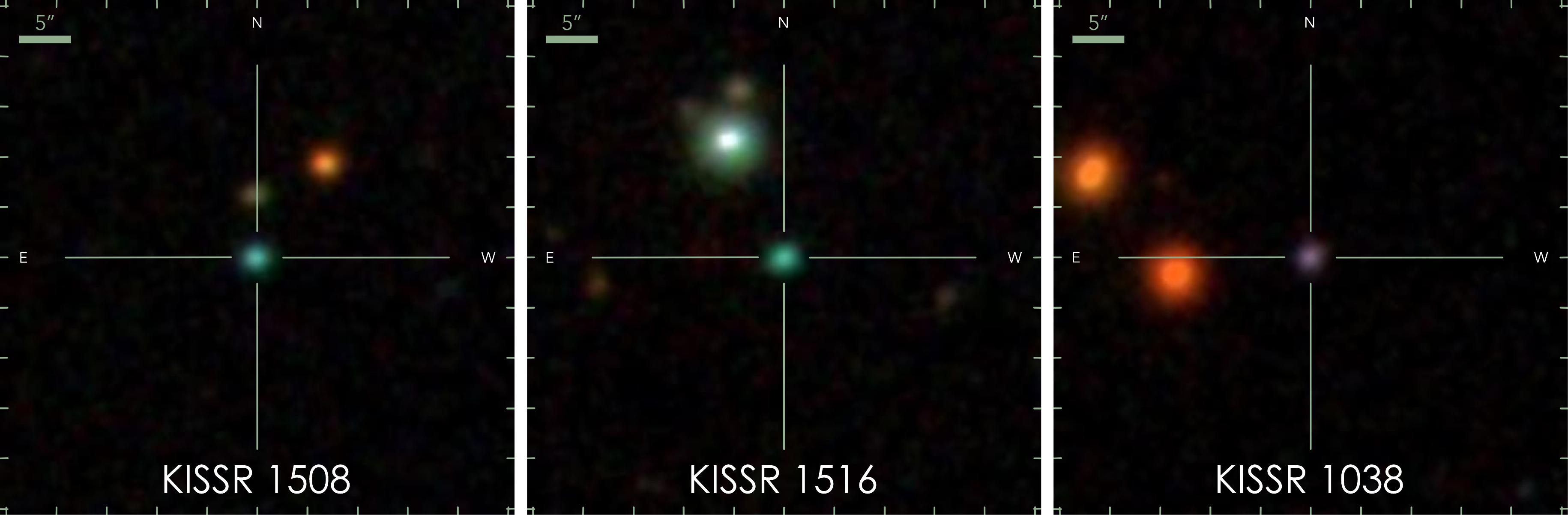}
\caption{\footnotesize SDSS images of three of our [O~III]-detected KISSR galaxies.  Both KISSR 1508 and 1516 exhibit the green color for which the Green Pea galaxies are known.  These two ELGs have redshifts of 0.294 and 0.328, respectively, which places them in the same redshift range as the \citet{Cardamone2009} GPs (see Figure~\ref{fig:zHist}).  However, KISSR 1038 (z = 0.410) appears violet in the SDSS three-color image because its strong [O~III] line is redshifted out of the SDSS r filter and into the SDSS i filter (which maps to red in the SDSS RGB color images).  Hence higher redshift Green Peas transition into Purple Grapes.
}
\label{fig:images}
\end{figure*}

\subsection{Properties of the [O~III]-detected KISSR Galaxies}
We summarize the observed and derived characteristics of the KISS [O~III]-detected star-forming galaxies in Table~\ref{tab:prop}.  Column 1 lists the KISSR number, while columns 2 and 3 give the equatorial coordinates of our galaxies (J2000).  We note in passing that the declination for KISSR 847 published in the original survey paper \citep{salzer2001} was incorrect; it corresponds to an r=17.15 star located 13.6 arcsec due N of KISSR 847.  The coordinates listed here are the correct ones.  Columns 4 and 5 list the B magnitude and B$-$V color for each galaxy, corrected for Galactic absorption and reddening.  The measured redshifts and [O~III] equivalent widths, derived from the KISS quick-look spectra, are given in columns 6 and 7.  Formal uncertainties in the KISSR photometric and spectroscopic quantities listed here are included in the survey papers cited above.

B-band absolute magnitudes are listed in column 8.  These are observed, as opposed to rest-farme (i.e., K corrected), luminosities; see the discussion in \S 4.  Typical uncertainties are 0.1-0.2 mag.  The median value of M$_B$ = $-$21.02 for KISSR 1508 is somewhat brighter than M$_B^*$ for the local B-band luminosity function \citep[approximately $-$20.0;][]{CFA1988, efstathiou1988}.   However, the stellar masses listed in column 9 are more modest in scale.  The median stellar mass of M$_*$ = 10$^{9.55}$ M$_\odot$, also for KISSR 1508, is lower than the characteristic mass determined from the stellar mass function \citep[1-2 $\times$ 10$^{11}$ M$_\odot$;][]{bell2003b,panter2004}.   Oxygen abundances are listed in column 10.  These values are derived using the O3N2 strong-line method, and have formal uncertainties of 0.10-0.15 dex \citep{hirschauer2018}.  Improved metallicities obtained using the direct method will be presented in Brunker \etal (in preparation).  Finally, column 11 presents star-formation rates (SFR).  These are derived using the H$\alpha$ line fluxes and employ the standard \citet{kennicutt} SFR conversion factor (SFR = L$_{H\alpha}$/7.9$\times$10$^{42}$).  Typical uncertainties in the SFR values are $\sim$10-20\%, where the error is dominated by the uncertainty in the H$\alpha$ flux measurement.

The derived properties of the KISSR [O~III]-detected galaxies presented in Table~\ref{tab:prop} will be compared with similar quantities from samples of GP galaxies and local star-forming galaxies in the following sections.

\section{KISSR [O~III]-detected galaxies as Green Peas} \label{sec:KISSasGP}

The KISSR [O~III]-detected star-forming galaxies were cataloged using a very different selection method than that used in creating the GP galaxy sample \citep{Cardamone2009}, yet the spectra of the two sets of galaxies appear very similar.  In the current section, we compare observed and derived quantities for the two samples of galaxies, with the goal of determining the extent to which the two groups can be thought of as forming one class of galaxies.  The data for the Green Peas come from \citet{Cardamone2009}, except for the oxygen abundances which come from \citet{Izotov2011}

\begin{figure}[t]
\epsscale{1.2}
\plotone{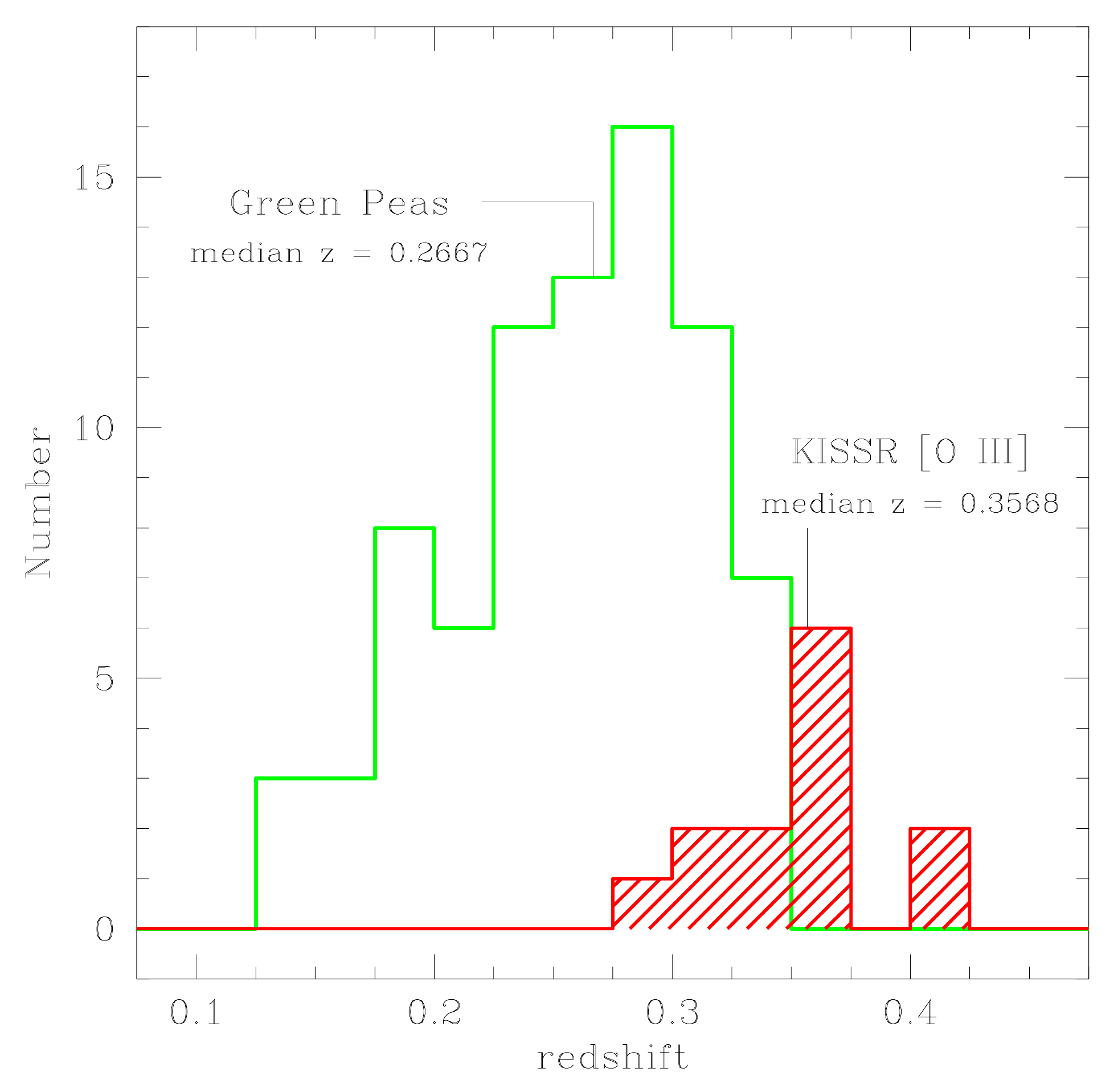}
\caption{\footnotesize Redshift histograms for the \citet{Cardamone2009} Green Pea (GP) galaxies (green histogram) and the KISSR [O~III]-detected star-forming galaxies (red shaded histogram).  The redshift ranges of the two samples are strongly defined by the selection methods used to detect the galaxies, as descibed in the text.  The median KISSR redshift is slightly larger than the maximum redshift included among the \citet{Cardamone2009} GP galaxies (z = 0.348). }
\label{fig:zHist}
\end{figure}

It is well understood that the observational characteristic that makes the GP galaxies green in the SDSS color images is the presence of a very strong [O~III]$\lambda$5007 emission line that redshifts into the SDSS r-band filter (r is mapped to green in the SDSS color images).  The redshift range of the \citet{Cardamone2009} GP galaxies is 0.141 $<$ z $<$ 0.348 (see Figure~\ref{fig:zHist}).  These redshifts nicely bracket the 50\% throughput range of the SDSS r-band filter for the location of redshifted  [O~III]$\lambda$5007 (5710 -- 6750 \AA).  At higher redshifts, the [O~III] line is shifted out of the r-band into the i-band, so these objects no longer appear green in the color images and are not selected as GPs.   Analogously, the redshift range of the [O~III]-detected KISSR galaxies (0.294 $\leq$ z $\leq$ 0.420) is limited by the KISS Red filter, which had a wavelength coverage of $\sim$6400 -- 7200 \AA\   \citep{salzer2000}.  

The impact of the combination of line strength and redshift is illustrated in Figure~\ref{fig:images}, which shows SDSS three-color images for three of the KISSR [O~III]-detected galaxies.  The two lower redshift systems, KISSR 1508 and KISSR 1516 (whose spectrum is shown in Figure~\ref{fig:spec}), exhibit the characteristic green color seen in the \citet{Cardamone2009} GP galaxies.  Both of these galaxies have redshifts that place the strong [O~III] line in the SDSS r-band filter.  However, the third galaxy shown, KISSR 1038, has a redshift that places the [O~III] line in the SDSS i-band filter, which maps to red in the SDSS images.  Presumably it is the strong line emission in the SDSS i filter combined with strong blue continuum emission in the SDSS g filter (mapped to blue) that gives rise to the strong purple hue seen in this galaxy.

\begin{figure}[ht]
\epsscale{1.2}
\plotone{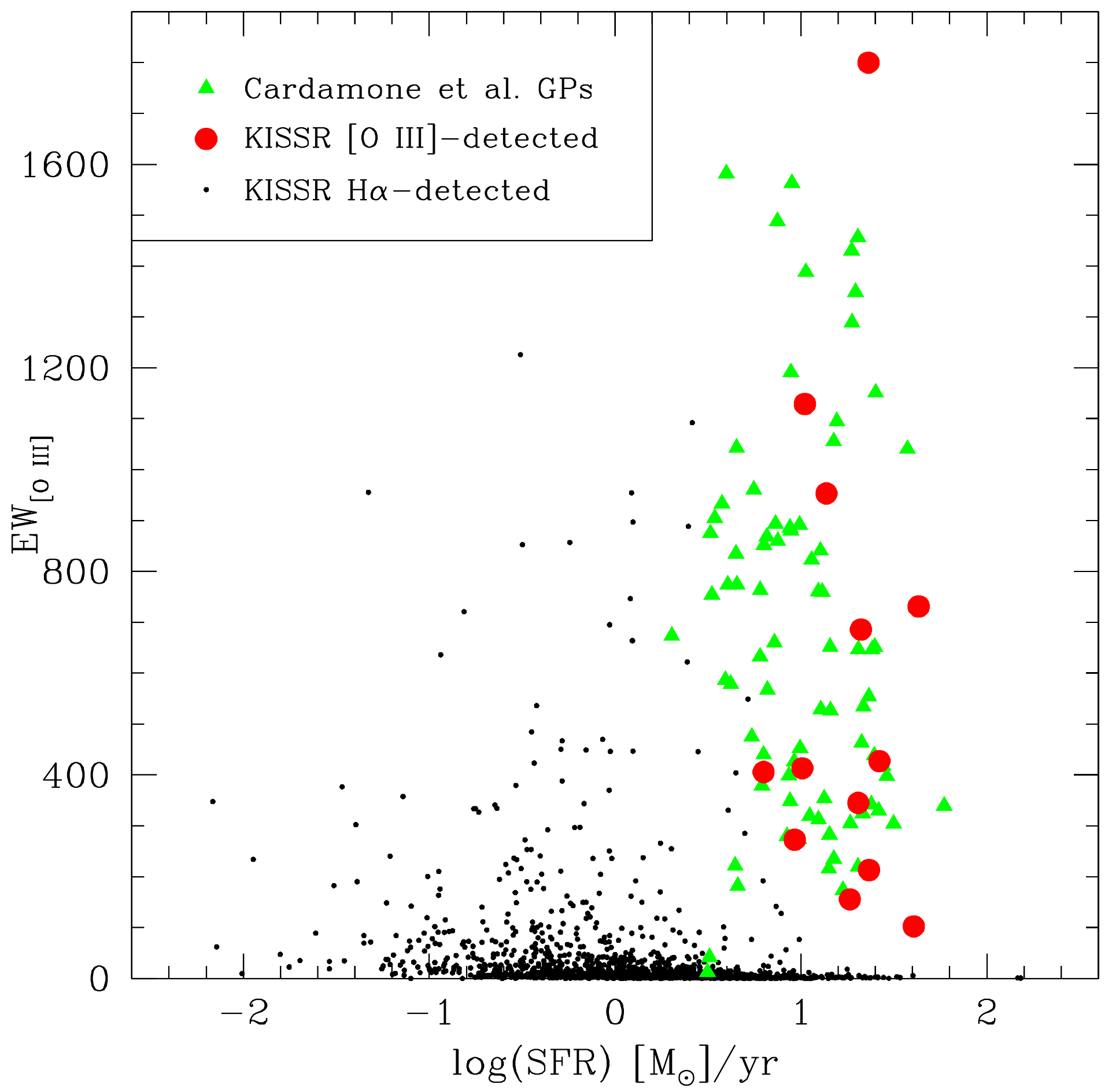}
\caption{\footnotesize Observed [O~III]$\lambda$5007 equivalent width plotted against the star-formation rate (SFR; derived from H$\alpha$ line fluxes) for the \citet{Cardamone2009} GP galaxies (green dots), the KISSR [O~III]-detected star-forming galaxies (red dots), and the KISSR H$\alpha$-detected star-forming galaxies (small black dots).  The KISSR [O~III]-detected galaxies occupy the same region as the GP galaxies: high SFR and large [O~III] equivalent widths.  There is surprisingly very little overlap with the local KISSR H$\alpha$-detected sample, despite the fact that this survey goes out to z = 0.095 and detects the strongest star-forming galaxies within the local volume.}
\label{fig:EWvSFR}
\end{figure}

The comparison of the redshifts of the two samples shown in Figure~\ref{fig:zHist} reveals that there is substantial overlap. The KISSR [O~III]-detected sample extends to somewhat larger redshifts, with the median redshift of 0.3568 lying just beyond the maximum redshift in the GP sample.  This leads to the expectation that the KISSR [O~III]-detected galaxies might be systematically higher luminosity systems.  As we see below, this expectation is at least partially borne out.

\begin{figure}[ht]
\epsscale{1.2}
\plotone{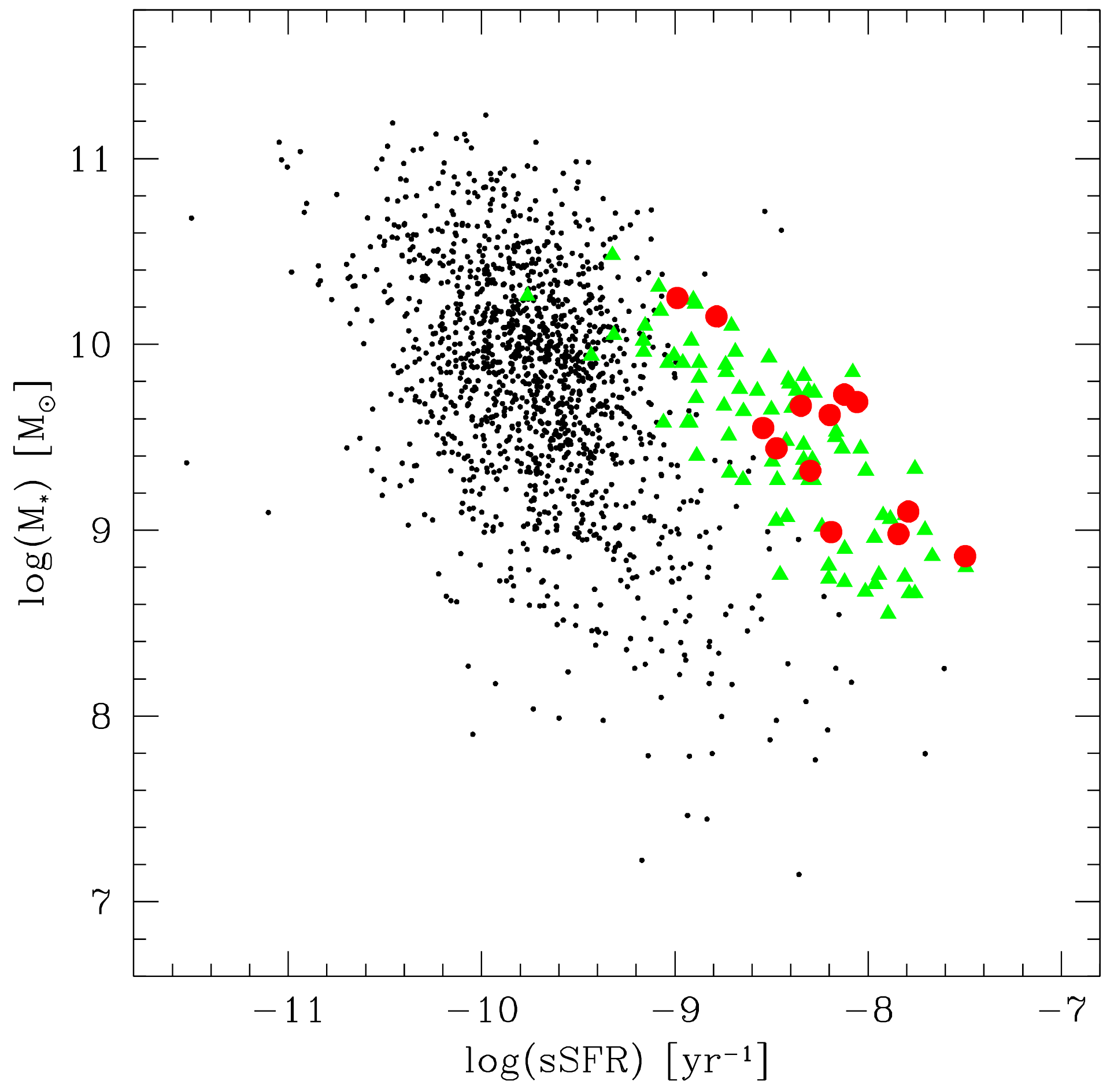}
\caption{\footnotesize Stellar mass plotted vs. the specific star-formation rate (sSFR).  The symbols have the same meaning as those in Figure~\ref{fig:EWvSFR}.  The GP and [O~III]-detected KISSR galaxies occupy a broad range of intermediate masses.  They are quite distinct from the nearby H$\alpha$-selected star-forming galaxies, having extremely large sSFR values.}
\label{fig:MvsSFR}
\end{figure}

While the specific filters being used to create the two samples dictate the range of observed redshifts that can be detected, the key parameter that determines whether a galaxy will be selected as a GP or a KISS ELG is the emission line strength.  For both samples it is a combination of the [O~III]$\lambda$5007 line flux and the line equivalent width that controls whether or not a source will be selected.  Figure~\ref{fig:EWvSFR} shows the equivalent width of the [O~III]$\lambda$5007 line (EW$_{[O~III]}$) vs. the logarithm of the star-formation rate (SFR) for the KISSR [O~III]-detected star-forming galaxies (red circles) and the \citet{Cardamone2009} Green Peas (green circles).  The small black points are the lower redshift KISSR H$\alpha$-detected star-forming galaxies.  In all cases the SFR is derived from the measured H$\alpha$ line luminosity assuming the \citet{kennicutt} conversion factor.

We highlight two important results illustrated in Figure~\ref{fig:EWvSFR}.  First, the KISSR [O~III]-detected galaxies occupy the same region of the diagram as the \citet{Cardamone2009} Green Peas.  This supports the hypothesis that the KISSR [O~III]-detected galaxies are drawn from the same population of objects as the GPs.  The KISSR [O~III]-detected objects cover the same range of EW$_{[O~III]}$, but tend to lie at the higher values of SFR (i.e., higher H$\alpha$ luminosities).  As alluded to above, the fact that the KISSR [O~III]-detected galaxies are located at greater distances than the GPs means that they will preferentially sample the higher luminosities.  Many of the \citet{Cardamone2009} Green Peas have SFRs lower than any of the [O~III]-detected KISSR galaxies, but these tend to be located at smaller distances not covered by the KISSR [O~III]-detected sample. 

\begin{figure}[ht]
\epsscale{1.2}
\plotone{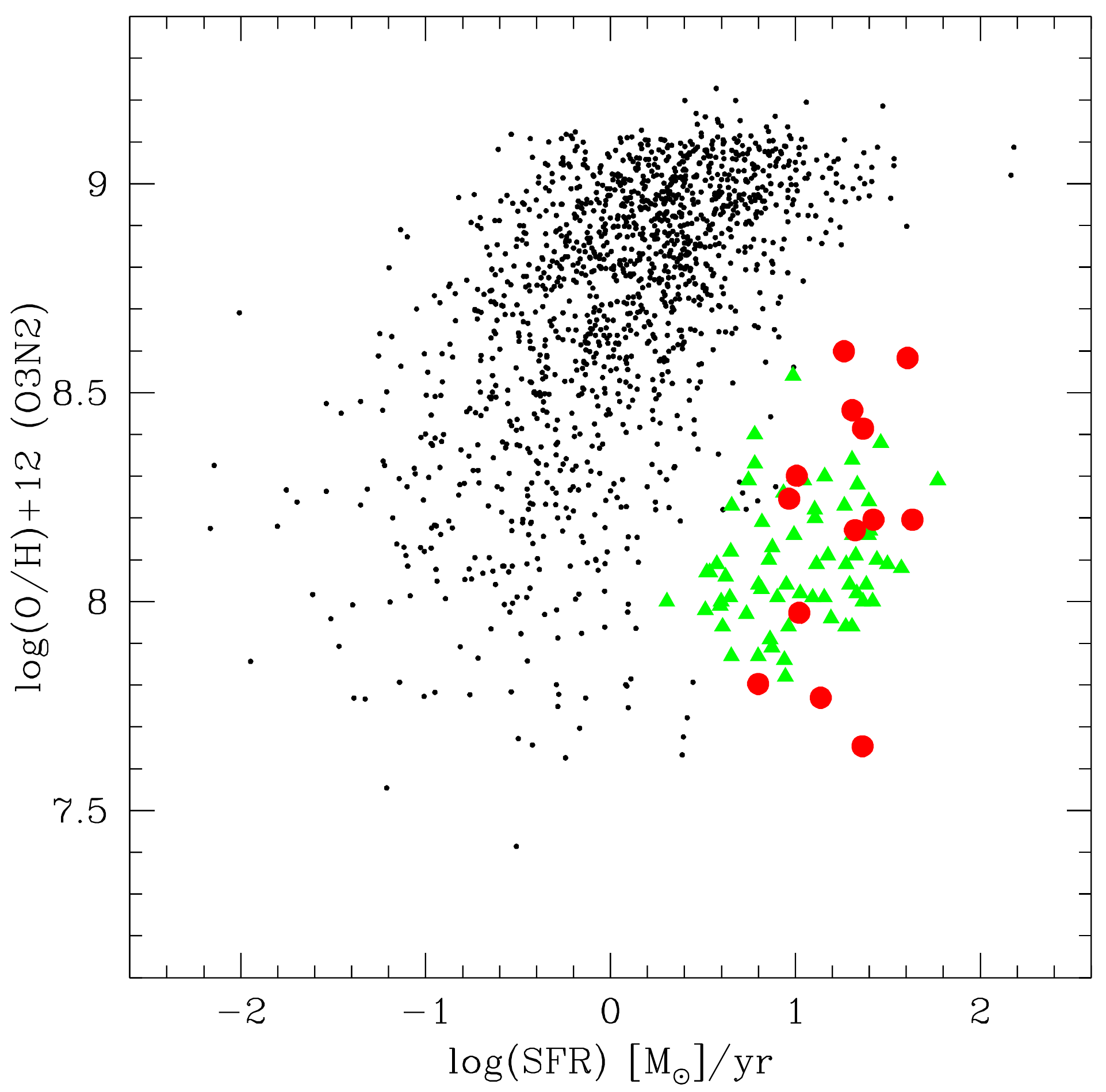}
\caption{\footnotesize Oxygen abundance plotted against SFR. The symbols have the same meaning as those in Figure~\ref{fig:EWvSFR}.  As in the previous figures, the GP and [O~III]-detected KISSR galaxies show almost no overlap with the local star-forming galaxies. }
\label{fig:AbvSFR}
\end{figure}

The second point to stress about the distributions shown in Figure~\ref{fig:EWvSFR} is the unique nature of the the Green Pea galaxies.  The small black dots represent $\sim$1390 H$\alpha$-detected star-forming galaxies in the local Universe.  They are from the deep, flux-limited KISSR sample.  Yet there is virtually no overlap between the H$\alpha$-detected KISSR galaxies and the GPs.  There are a number of KISSR galaxies with SFRs in the same range as the GPs, but none have EW$_{[O~III]}$ values comparable to the GPs.   Likewise, there are many H$\alpha$-detected KISSR galaxies with large equivalent widths, but these all have lower SFRs.  These are primarily star-forming dwarfs.   Only a handful of the H$\alpha$-detected KISSR galaxies are located among the GPs in Figure~\ref{fig:EWvSFR}, and these few are on the periphery of the distribution.  Despite the fact that the H$\alpha$-detected portion of the KISSR sample have redshifts out to z = 0.095, there are no objects within the survey volume that have properties as extreme as the majority of the GP galaxies.  We return to this point in the following sections.

Next we plot the logarithm of the stellar mass (M$_*$) vs. the logarithm of the specific star-formation rate (SFR normalized by M$_*$) for the same three samples of star-forming galaxies in Figure~\ref{fig:MvsSFR}.  The same two trends we highlighted in Figure~\ref{fig:EWvSFR} are seen here.  The \citet{Cardamone2009} GPs and the KISSR [O~III]-detected galaxies are co-located in the diagram, where again the KISSR [O~III]-detected galaxies are trending toward the extremes of the distribution.  Once again, the GP galaxies are located in a region of the diagram that is largely devoid of examples from the low-redshift Universe.  Compared to Figure~\ref{fig:EWvSFR}, there is somewhat more overlap between the samples in this figure, but nonetheless the GPs are clearly seen to be extreme.  The H$\alpha$-detected KISSR galaxies with sSFR values comparable to the GPs are lower mass galaxies.

It is worth noting here that the GPs and KISSR GP-like galaxies possess a range of stellar mass values that span nearly two orders of magnitude, from 10$^{8.5}$ to 10$^{10.5}$ M$_\odot$.  While some authors have referred to the population of Green Pea galaxies as being dwarfs, it is clear that they actually span the full extent of intermediate masses, and that the largest examples are quite massive.

Our last comparison of the KISSR samples with the \citet{Cardamone2009} Green Peas involves oxygen abundance and SFR.  Figure~\ref{fig:AbvSFR} shows similar trends to the previous two plots.  The GPs and KISSR [O~III]-detected galaxies once again occupy the same region of parameter space.  They are also almost completely offset from the KISS H$\alpha$-selected sample towards higher SFRs and/or lower abundances.  

Figure~\ref{fig:AbvSFR} hints at another important characteristic of Green Pea and GP-like galaxies: they are metal poor.  When compared to galaxies with comparable SFRs, the GP galaxies exhibit, on average, nearly a full dex lower metallicities. We will return to this key feature in the next section.

Figures \ref{fig:zHist}$-$\ref{fig:AbvSFR} all tell a consistent story: the \citet{Cardamone2009} Green Peas and the KISSR [O~III]-detected star-forming galaxies have very similar properties.  We conclude that the KISSR [O~III]-detected galaxies are consistent with being drawn from the same population of extremely compact, high SFR systems as the ``classic" Green Peas.  From this point on, we will refer to the KISSR [O~III]-detected star-forming galaxies as the KISSR Green Peas. 

\section{Discussion: The Nature and Evolutionary Status of the KISSR Green Peas}
\label{sec:EvolutionaryStatus}

Having established the similarities of the \citet{Cardamone2009} Green Peas and the KISSR [O~III]-detected star-forming galaxies, we next switch our focus to trying to understand the nature of the KISSR GPs.  Unfortunately, their substantial distances render studies involving morphology and physical size using our available ground-based data useless: the KISSR GPs are all unresolved in available images.  Clearly, high-resolution space-based imaging would be highly useful in helping to understand the nature of these enigmatic objects.   Nonetheless, we can make substantial progress in understanding these systems using the available data. This comparative study is aided by the availability of the large sample of lower redshift H$\alpha$-detected KISSR galaxies.

\begin{figure}[ht]
\epsscale{1.2}
\plotone{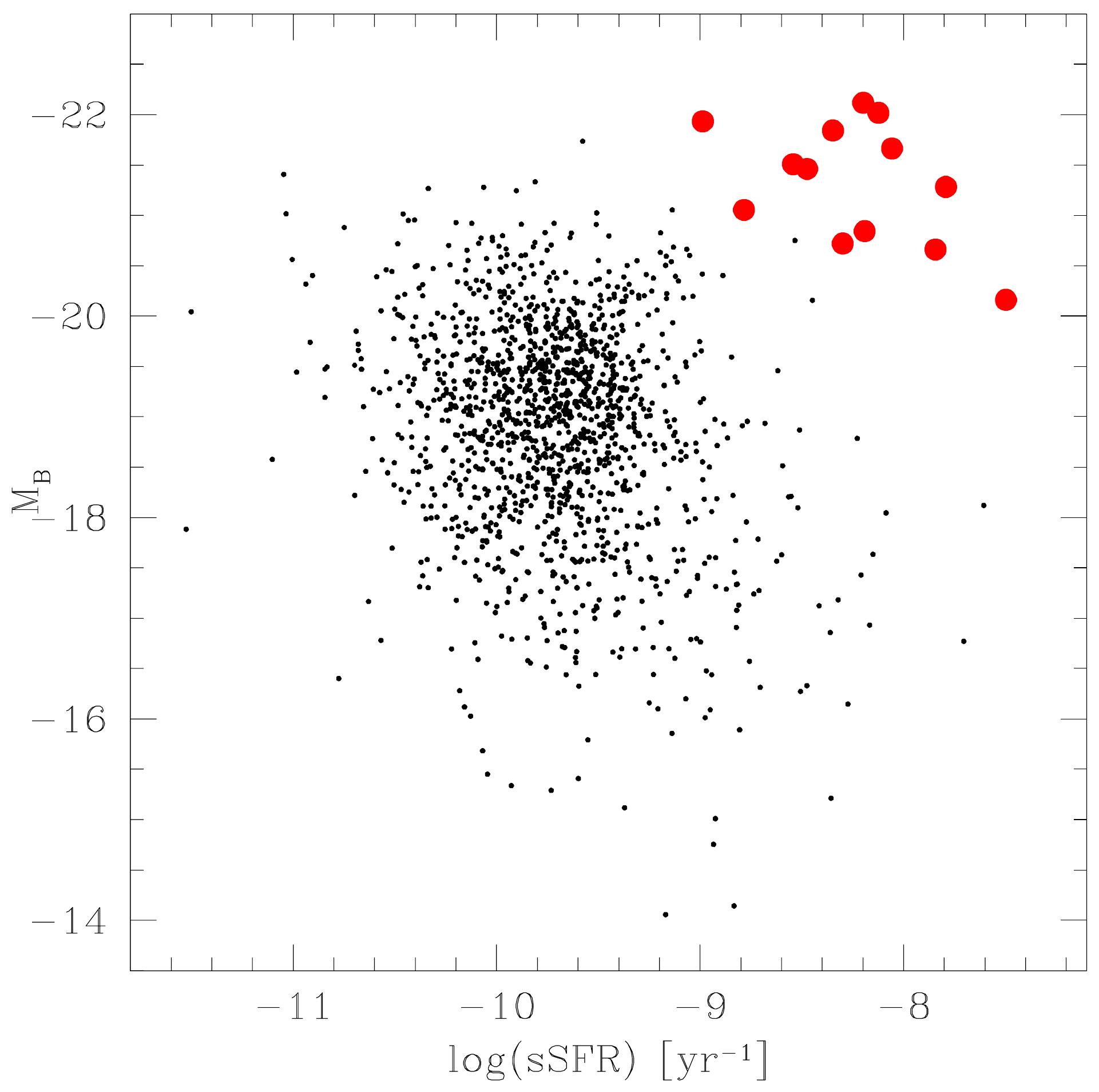}
\caption{\footnotesize B-band absolute magnitude plotted against the specific star-formation rate (sSFR) for the KISSR [O~III]-detected star-forming galaxies (red dots) and the KISSR H$\alpha$-detected star-forming galaxies (small black dots). }
\label{fig:kiss_ssfr_mb}
\end{figure}

We begin our efforts to place the KISSR GPs into context by looking at their optical luminosities.  Figure~\ref{fig:kiss_ssfr_mb} plots B-band absolute magnitude versus log(sSFR) for the KISS Green Peas (red circles) and the KISSR H$\alpha$-detected galaxies (black points).   We note in passing that the \citet{Cardamone2009} study did not include an analysis of the optical luminosities of their Green Pea sample, nor did they tabulate absolute magnitudes for their GPs.  Hence, the \citet{Cardamone2009} GPs will not appear in the figures presented in this section of the paper.

In order to account for the redshift differences between the two samples, one would typically impose a ``K correction" \citep{pence1976, hogg2002} to the light of one or both samples.  However, the strongly star-bursting nature of the KISSR GPs makes the implementation of {\it standard} K corrections inadequate.  Rather, we need a correction that reflects the actual emission spectrum of these extreme objects.  We get around this difficulty by recognizing that the redshifts of the two KISSR samples are such that the wavelength coverage of the H$\alpha$-detected KISSR galaxies in the B-band is comparable to the wavelength coverage of the [O~III]-detected KISSR galaxies as covered by the V-band.  In other words, we can simply substitute M$_V$ for the KISSR GP galaxies for the comparison with M$_B$ for the H$\alpha$-detected KISSR galaxies.  This substitution is not perfect, but to first order accounts for the redshift differences between the two samples.   In Figure~\ref{fig:kiss_ssfr_mb} and all subsequent figures utilizing B-band absolute magnitudes, the quantity being plotted for the KISSR GP galaxies is actually the {\it measured} V-band absolute magnitude M$_V$.  

\begin{figure*}[ht]
\centering
\includegraphics[width=0.49\linewidth]{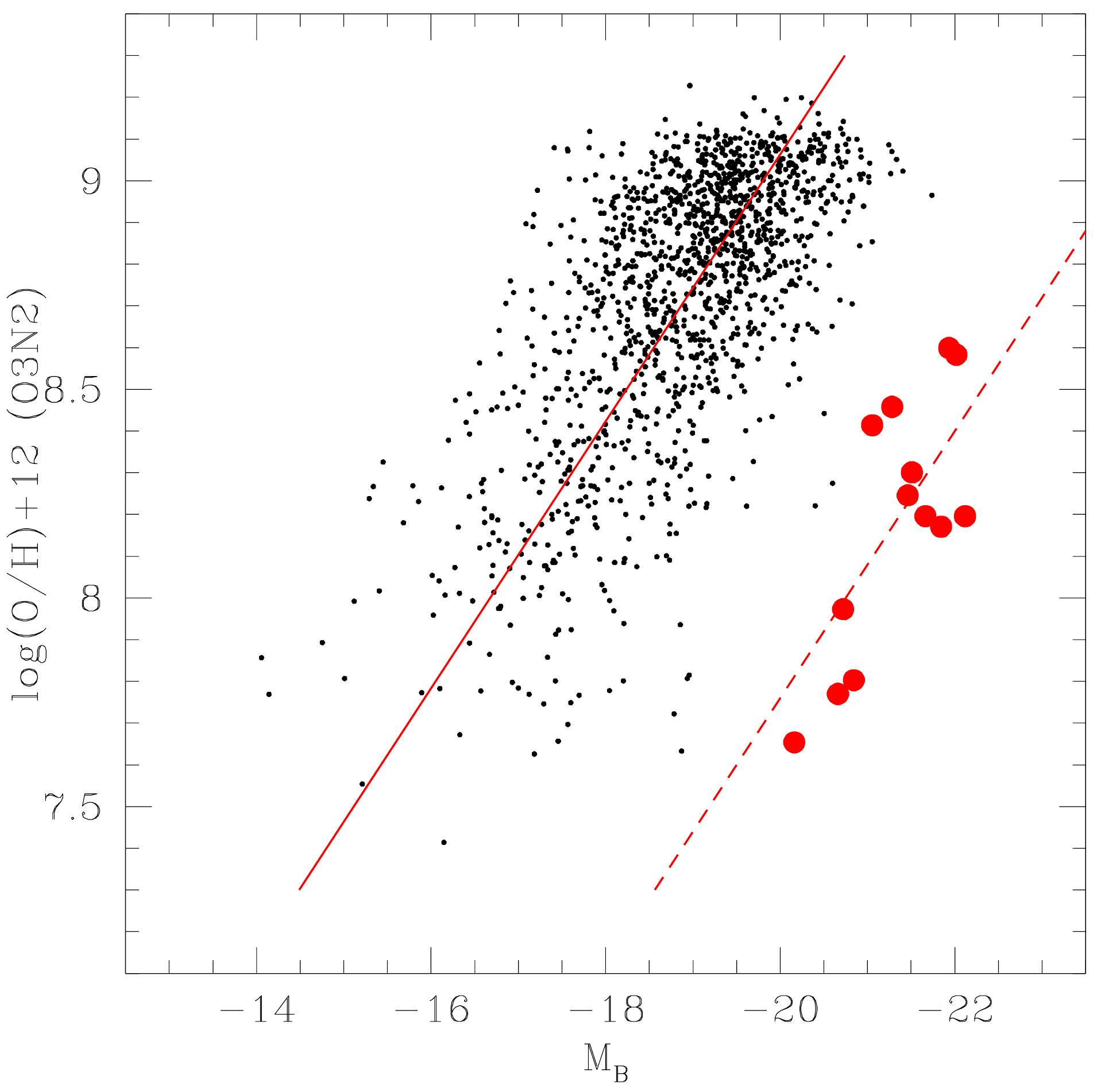}\quad\includegraphics[width=0.49\linewidth]{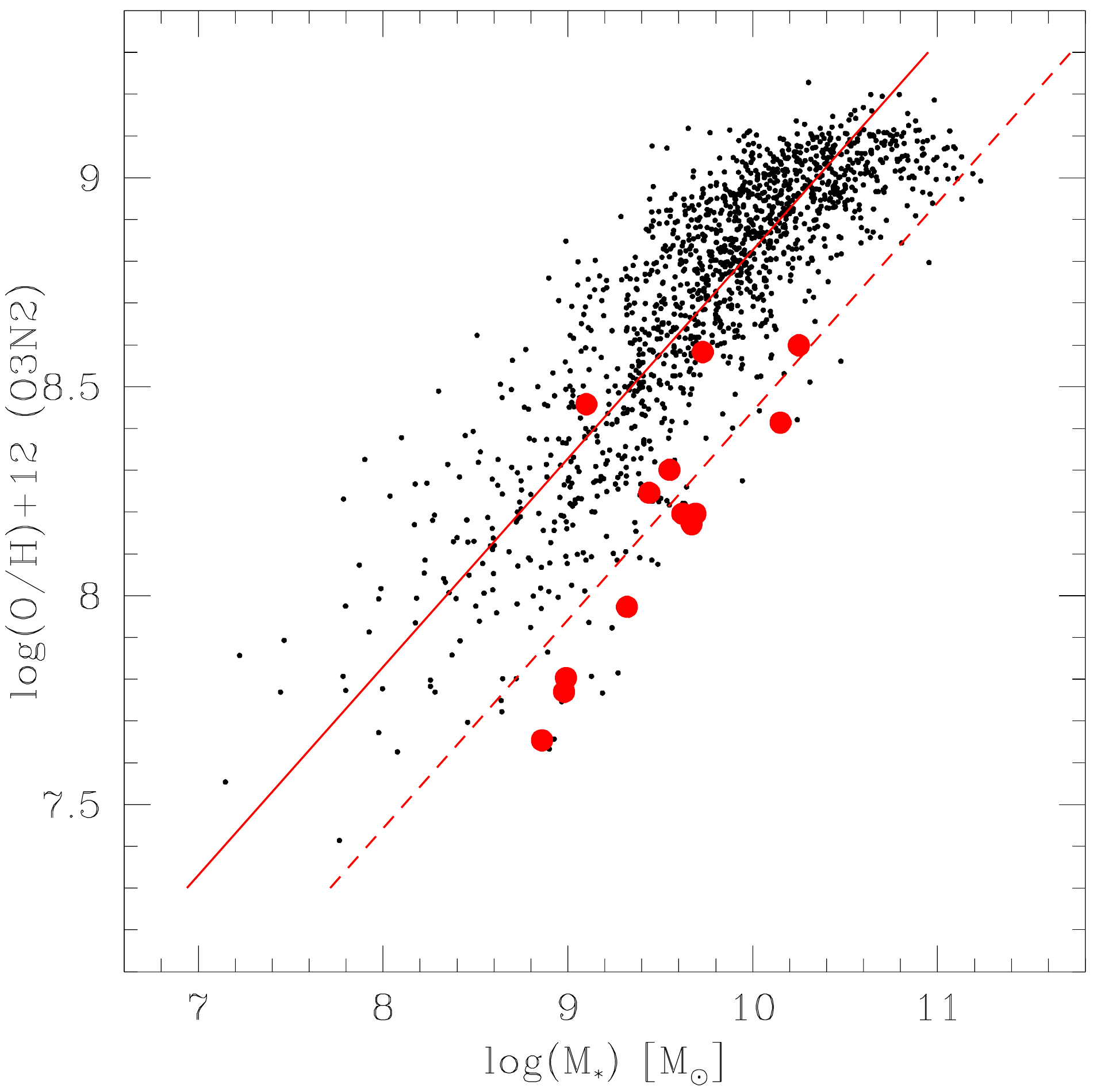}
\caption{\footnotesize Luminosity-Metallicity (LZ) diagram (left) and Mass-Metallicity (MZ) diagram (right) for the KISSR galaxies.  The symbols have the same meanings as those in Figure~\ref{fig:kiss_ssfr_mb}.  In both figures, the solid red line is a bivariate linear least squares fit to the low-redshift KISSR galaxies, adopted from \citet{hirschauer2018}.  The dashed red lines are fits to the KISSR GPs, constrained to have the same slope as the fit to the low-z galaxies.  The LZ diagram is similar to the one presented in \citet{Salzer2009}, but uses updated abundances.  The [O~III]-selected galaxies are offset by a huge amount from the local star-forming galaxies.  In the MZ diagram the offset is still substantial but is much reduced.  This shows that the offset seen in the LZ plot is a combination of both abundance and luminosity offsets.}
\label{fig:mz_lz_kiss}
\end{figure*}

We stress that this is not a true K correction, since the corrected KISSR GP's absolute magnitudes are not "rest frame" luminosities, but rather are shifted to match the average spectral bandpass of the KISSR H$\alpha$-detected galaxies.  As long as the comparison is strictly between these two samples, however, the correction should be valid.

As seen in Figure~\ref{fig:kiss_ssfr_mb}, the KISS Green Peas are very luminous compared to the average galaxy in the KISS H$\alpha$-selected sample.  The median (corrected) B-band absolute magnitude of the KISSR GPs is $-$21.5, roughly 1.5 magnitudes above M$^*_B$ for local galaxy samples \citep{CFA1988, efstathiou1988}).   As we saw previously, the KISS Green Peas have much larger values of log(sSFR) at a given mass/luminosity compared with the KISS H$\alpha$-selected sample.  However, comparison with Figure~\ref{fig:MvsSFR} shows that the offset between the KISSR Green Peas and the KISSR H$\alpha$-detected sample is much more extreme when we substitute optical luminosity for mass.  While the KISSR GPs lie in the middle of the range of stellar masses for the local KISSR galaxies, they are located at the very extreme end of the luminosity distribution.  This is a strong indication of the strength of the starburst on the Green Pea galaxies.

Important clues about the nature and evolutionary status of the KISSR GPs galaxies can be inferred from their measured metallicities.  All previous studies of GP-like galaxies have found that these objects have low oxygen abundances \citep[e.g.,][]{Cardamone2009, Izotov2011}. 
\citet{Salzer2009} presented a luminosity-metallicity (LZ) plot that showed that the [O~III]-detected KISSR galaxies were located more than one dex below the LZ trend defined by the nearby H$\alpha$-detected KISSR galaxies.  In that work the offset between the two galaxy samples was interpreted as being primarily a metallicity offset.  Here we show that this interpretation was too simplistic.

We present LZ and mass-metallicity (MZ) relations that compare the two galaxy samples in Figure~\ref{fig:mz_lz_kiss}.  In both figures the solid red line represents a bivariate linear least-squares fit to the low-reshift H$\alpha$-detected KISSR galaxies, taken directly from \citet{hirschauer2018}.  The dashed red lines are fits to the [O~III]-detected KISSR galaxies, constrained to have the same slope as the fits to the H$\alpha$-selected galaxies.  In the LZ-relation shown in the left-hand panel, the KISSR Green Peas do not overlap the KISSR H$\alpha$-detected sample at all.  
The vertical (metallicity) offset between the two fit lines is 1.30 dex, while the horizontal offset is 4.07 magnitudes!  We demonstrate below that the dramatic offset between the two samples is caused by a combination of an inherent metallicity offset and a strong luminosity enhancement due to the starburst.

\begin{figure}[ht]
\epsscale{1.2}
\plotone{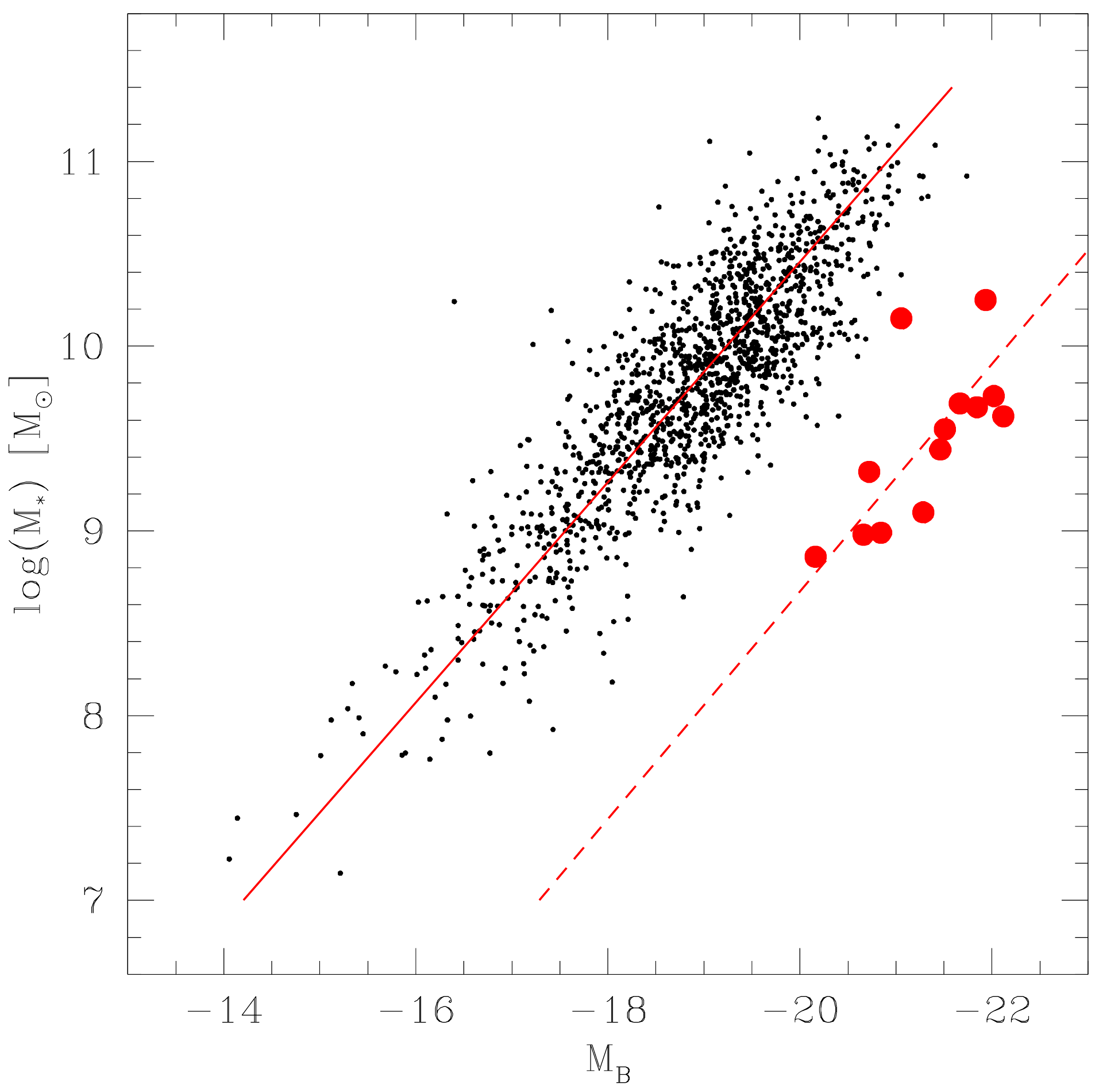}
\caption{\footnotesize Stellar mass plotted against B-band absolute magnitude for the KISSR galaxies.  The symbols have the same meanings as those in Figure~\ref{fig:kiss_ssfr_mb}.  The solid red line is a bivariate linear least squares fit to the low-redshift KISSR galaxies, while the dashed red line is a fit to the KISSR GPs, constrained to have the same slope as the fit to the low-z galaxies.  The luminosity enhancement for the [O~III]-detected KISSR galaxies at a given mass is, on average, 3.08 magnitudes (factor of 17).}
\label{fig:mass_mb_kiss}
\end{figure}

 The MZ relations are plotted in the right-hand side of Figure~\ref{fig:mz_lz_kiss}.   Now the offset of the KISSR Green Peas from the KISSR H$\alpha$-detected sample is substantially reduced.  By replacing luminosity with stellar mass on the horizontal axis, the luminosity enhancement due to the starburst is effectively removed.  We argue that the remaining offset seen in the MZ plot is the intrinsic metallicity offset of the GPs relative to the nearby star-forming galaxies.  The vertical offset between the two fits  implies that the KISSR GPs are, on average, 0.39 dex lower metallicity.   A few of the KISSR GPs (KISSR 225 and 2005) straddle the fit line for the low-redshift sample and are not metal-deficient at all.  The remaining KISSR GPs fall within the lower envelope of galaxies in the MZ relation.  They are clearly metal-poor for their masses, but not as extreme as their appearance in the LZ diagram seemed to indicate.

A simple but effective way to quantify the strength of the starbursts on the KISSR GP galaxies is presented in Figure~\ref{fig:mass_mb_kiss}.  Here we plot stellar mass against luminosity for both KISSR samples.  In this figure the horizontal offset between the two sets of galaxies directly measures the strength of the starburst in the Green Pea galaxies.  Based on the horizontal offset between the two fit lines (which are determined in the same way as the fits in Figure~\ref{fig:mz_lz_kiss}), we find that the average luminosity enhancement due to the starburst events in the GP galaxies is 3.08 magnitudes (factor of 17 higher luminosity).  We stress that this offset is measured relative to a population of actively star-forming galaxies whose own luminosities are enhanced by some amount.  If measured against a population of quiescent galaxies the starburst enhancement is likely to be even higher than measured here.
It is also clear that the KISSR GPs exhibit a range of luminosity enhancements, from as low as 1.57 magnitudes for KISSR 1791 (factor of 4 higher luminosity) to as high as 3.55 magnitudes for KISSR 2005 (factor of 26 higher luminosity).

In the local Universe it is rare to find star-forming galaxies where more than half of the total light of the system is coming from the starburst population \citep[e.g.,][]{janowiecki2014}.   Such objects are typically blue compact dwarf galaxies with low luminosities and high sSFR values.  There does not appear to be any examples in the nearby Universe with star-formation events as extreme as the Green Pea galaxies.  The average object in our sample is 17$\times$ more luminous in the B-band than a typical local star-forming galaxy of the same mass.  This means that, on average, at least 94\% of the B-band light from these systems must be coming from the starburst population, which in turn seems to imply that at least some of these systems are undergoing their first major episode of star formation.

Is it possible that the Green Pea galaxies represent newly formed (or forming) systems?  The current sample of GPs have lookback times of 3-4 Gyrs.  While the general picture of galaxy formation has most major galaxies forming at much higher redshifts, there is no reason why some of these intermediate-mass systems couldn't be coming together and having their first major star-formation event at these late times.

\begin{deluxetable*}{ccccccc}
\tabletypesize{\footnotesize}
\tablewidth{0pt}
\tablecaption{The Extreme Nature of the KISSR [O~III]-detected Galaxies \label{tab:extreme}}

\tablehead{
 \colhead{KISSR} &  \colhead{Corrected} &  \colhead{Starburst} &  \colhead{Fraction of Light} &  \colhead{Star-Formation} & \colhead{Birthrate} & \colhead{Burst Mass} \\
  &  \colhead{M$_B$} & \colhead{Enhancement} & \colhead{from Starburst}& \colhead{Age} & \colhead{Parameter b} & \colhead{Fraction} \\
  &  \colhead{} & \colhead{[mag]} & \colhead{[\%]} & \colhead{[Myr]} && \colhead{[\%]} \\ 
  (1)  & (2)  & (3)  & (4)  & (5) & (6) & (7)
}
 
\startdata
 \ 225 &  $-$22.02 &   3.24 &  94.9 &  133 & 24.2 & 52.3 \\
 \ 560 &  $-$20.84 &   3.30 &  95.2 &  155 & 73.6 & 52.8 \\
 \ 847 &  $-$20.16 &   2.84 &  92.7 &  \ 31 & 69.9 & 70.4 \\
  1038 &  $-$21.67 &   2.95 &  93.4 &  114 & 17.9 & \ 8.3 \\
  1290 &  $-$21.46 &   3.16 &  94.6 &  298 & 31.8 & 51.5 \\
  1508 &  $-$21.51 &   3.03 &  93.9 &  349 & 35.7 & 30.9 \\
  1516 &  $-$21.84 &   3.16 &  94.6 &  222 & 23.8 & 10.5 \\
  1759 &  $-$20.72 &   2.62 &  91.0 &  199 & 74.2 & 57.9 \\
  1791 &  $-$21.06 &   1.57 &  76.4 &  608 & \ 3.3 & \ 8.6 \\
  1825 &  $-$21.93 &   2.28 &  87.8 &  971 & \ 6.1 & \ 4.7 \\
  1953 &  $-$22.12 &   3.52 &  96.1 &  158 & 19.8 & 13.3 \\
  2005 &  $-$21.28 &   3.55 &  96.3 &  \ 62 & 113.7 & 56.1 \\
  2042 &  $-$20.66 &   3.13 &  94.4 &  \ 70 & 27.4 & 35.2  
\enddata
\end{deluxetable*}

An interesting parameter in this regard is the ``star-formation age" of these systems, obtained by dividing their stellar mass by their current SFR (i.e., just the inverse of the sSFR).  The star-formation age is the amount of time it would take to form the total stellar mass of the system at its current SFR.  This must be considered to be a lower limit to their true ages, since we do not necessarily expect their high SFR rates to be sustainable over long periods of time.  However, work by \citet{mcquinn2010} suggests that the lengths of starbursts in nearby low-mass galaxies can easily reach 500 Myr in duration.  Based on the numbers in Table~\ref{tab:prop}, the star-formation ages of several of the KISSR Green Peas are extremely low: 31 Myr for KISSR 847, 62 Myr for KISSR 2005, and 70 Myr for KISSR 2042.  In other words, these systems could form their entire mass of stars if the could sustain their current SFRs for several tens of millions of years.  The median star-formation age for our sample is 158 Myr, and 11 of the 13 galaxies have ages less than 350 Myr.  Based on these numbers, it certainly seems possible that at least some of the KISSR Green Peas were young systems at the lookback times were we observe them.

We list parameters in Table \ref{tab:extreme} that illustrate the extreme nature of the KISSR GP galaxies.  Column 2 gives the approximate K corrected B-band absolute magnitude (i.e., the observed V-band absolute magnitude), while column 3 lists the starburst enhancement from Figure \ref{fig:mass_mb_kiss}, defined as the horizontal offset of each GP galaxy from the trend line fit to the low-z KISSR galaxies.  Column 4 converts this enhancement into the fraction of the B-band luminosity coming from the starburst population (stellar plus nebular emission).  Column 5 lists the star-formation age as defined above.

While the values exhibited in Table \ref{tab:extreme} are indeed extreme, they are confirmed and supported by the results of our SED model fitting described in \S 2.2.  Even though our primary use of the CIGALE models for the current study has been to provide robust stellar mass determinations, the detailed SED fitting also generates estimates for several other key parameters \citep[see][for details]{janowiecki2017}.  Since the model results are based on the analysis of the full SED rather than on an observation in a single passband (e.g., B-band magnitude or H$\alpha$ line flux), they provide a largely independent measurement of the star-forming properties of the KISSR GPs.  Two relevant examples from these models are included in Table \ref{tab:extreme}.  

Column 6 of Table \ref{tab:extreme} lists the birthrate parameter b \citep{kennicutt1983}, defined as 
\begin{equation}
b = {SFR_{current}\over{\langle SFR\rangle}}.
\end{equation}
Here SFR$_{current}$ is the current star-formation rate, were we use the modeled value for the SFR over the past 10 Myr, and $\langle$SFR$\rangle$ is the lifetime average SFR.  The burst mass fraction is given in column 7.  This quantity is defined as the mass fraction of the young stellar population compared to the total stellar mass, were both are evaluated at the time of the initiation of the young population's formation burst.

The SED model results are in good general agreement with the observationally derived parameters listed in Table \ref{tab:extreme}, and paint a similarly extreme picture of the nature of the starburst episodes of the KISSR GPs.  A birthrate parameter b $\gtrsim$ 3 is generally adopted as an indicator that a strong starburst is occurring in a galaxy \citep[e.g.,][]{bergvall2016}.  All of the KISSR GPs have b $>$ 3, and in fact all but the two least severe systems (KISSR 1791 and 1825) have b $>$ 17.  Six of the KISSR GPs have burst mass fractions in excess of 50\%: more than half of the stellar mass in these systems has been produced in the current burst.  In the case of KISSR 847, the SED models indicate that more than 70\% of the stellar mass was produced during the initial stages of current starburst, which is broadly consistent with its extremely short star-formation age of 31 Myr.

It is worth noting here that the extremely short star-formation ages of the KISSR GPs is precisely one of the characteristics one looks for in a population of objects that could be contributing to the re-ionization of the Universe.  Current evidence suggests that re-ionization began at z $\sim$ 20, when the Universe was only $\sim$180 Myr old, and was largely complete by z $\sim$ 6 (Universe $\sim$950 Myr old).  The short formation timescales implied by the star-formation ages of the KISSR GPs would naturally allow the GP-like systems present in the early Universe to be major contributors to the ionizing photons that were driving the re-ionization.

\section{Number Densities of the [O~III]-Detected KISSR Galaxies}
\label{sec:VolumeDensities}

One of the strengths of the KISS sample of galaxies is that it represents a flux-limited catalog of emission-line galaxies with a well-defined selection function.  As detailed in \citet{gronwall2004b}, the completeness of KISS is well-parameterized by the emission-line fluxes measured from the digital objective-prism spectra.  Each survey strip has a well-defined line-flux completeness limit that allows us to derive accurate volume densities for the KISS ELGs.  This is in constrast to most existing samples of GP or GP-like galaxies \citep[e.g.,][]{Cardamone2009, Izotov2011}, where the samples were constructed based either on SDSS colors or by evaluation of existing SDSS spectra.

The line-flux completeness limit for the [O~III]-detected KISSR galaxies will be the same as was derived for the full KISSR sample \citep[e.g.,][]{gronwall2004b}, but the redshift range over which
the GP-like galaxies can be detected is substantially different from that of the KISSR H$\alpha$-detected sample.  Using the filter tracing of the KISS Red survey filter \citep[see][]{salzer2000}, we determine that the [O~III]$\lambda$5007 line would be detectable over the redshift range 0.290 to 0.422.  We adopt these redshift limits to determine the maximum volume over which the [O~III]-detected KISSR galaxies could be found.  When coupled with the sky coverage of the two catalogs (62.162 sq. deg. for KR1 and 65.864 sq. deg. for KR2) and using the cosmology parameters specified in \S 1, these redshift ranges yield effective co-moving volumes of 17.98 $\times$ 10$^6$ Mpc$^3$ for KR1 and 19.05 $\times$ 10$^6$ Mpc$^3$ for KR2.

Using the completeness limits for the KR1 and KR2 surveys together with the measured objective-prism line fluxes for the 13 KISSR [O~III]-detected galaxies, we derive a co-moving number density of 1.35 $\times$ 10$^{-6}$ Mpc$^{-3}$ for the KISSR GP galaxies.  This value supersedes a preliminary value for the density of these galaxies presented in \citet{Salzer2009}; the latter value is a factor of three smaller due to the simplifying assumptions used in its derivation.  

\begin{figure*}[ht]
\centering
\includegraphics[width=0.49\linewidth]{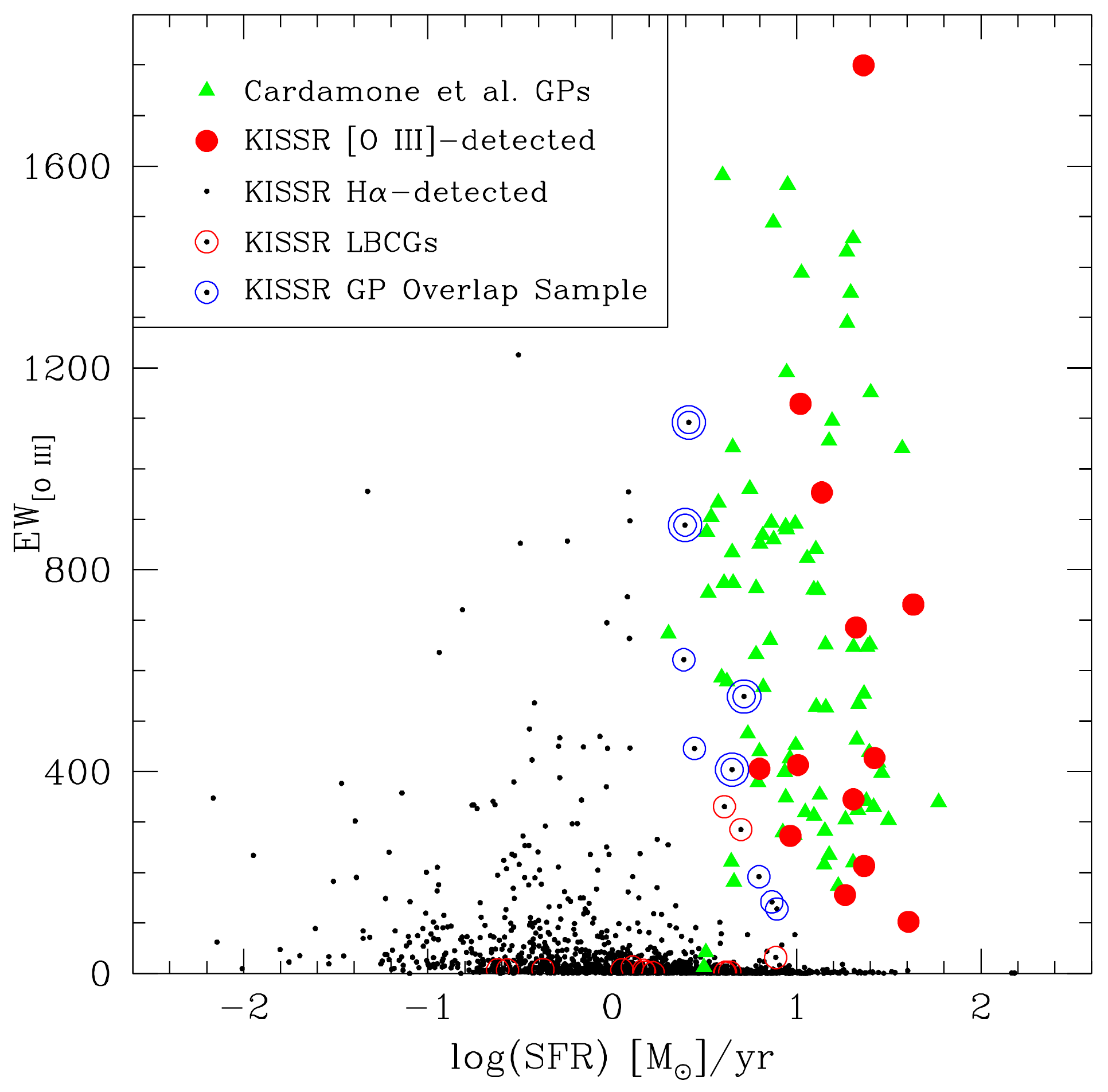}\quad\includegraphics[width=0.49\linewidth]{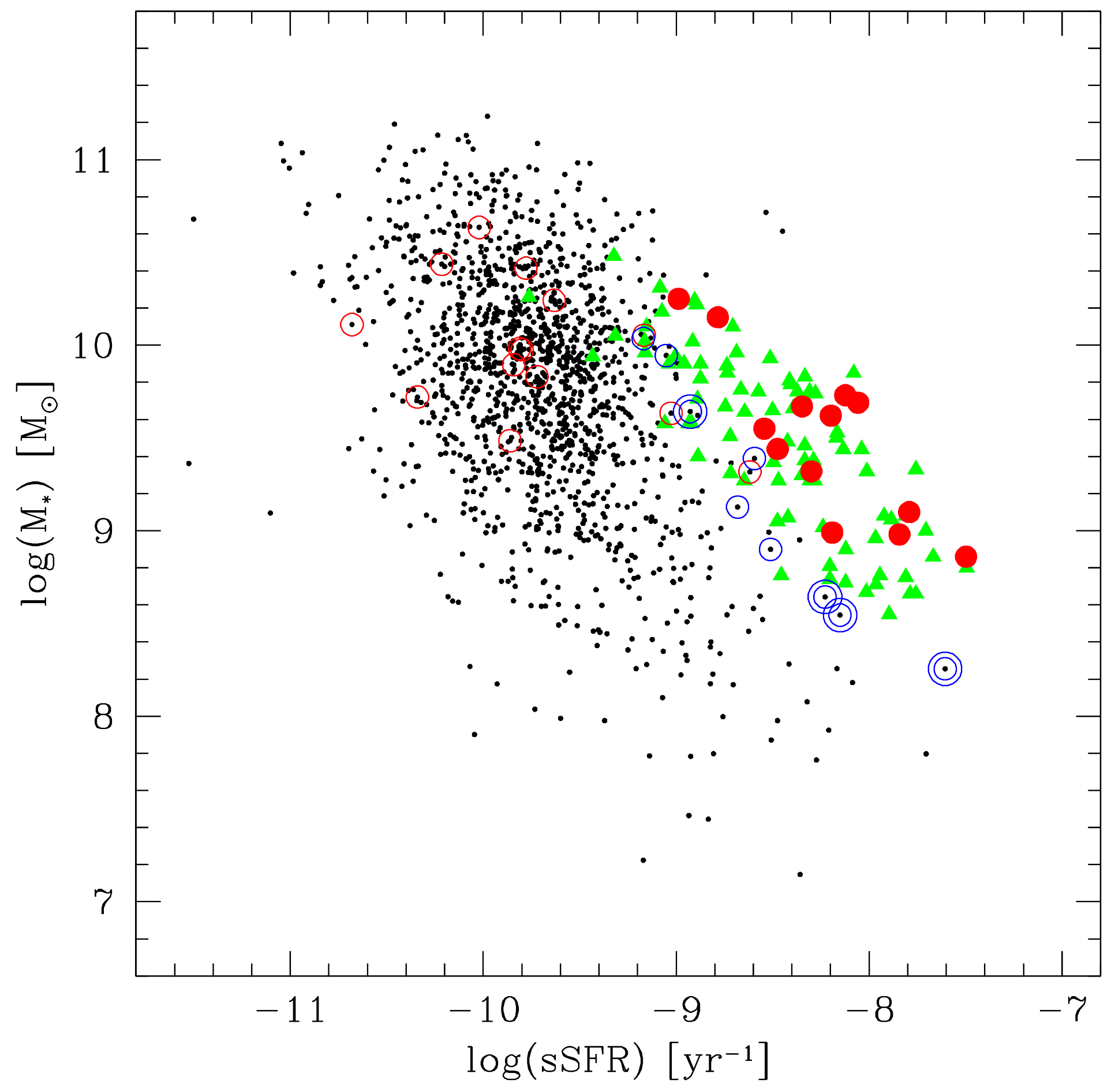}
\caption{\footnotesize We reproduce Figures \ref{fig:EWvSFR} and \ref{fig:MvsSFR} and indicate the two subsamples of KISS galaxies discussed in \S 6.1 and \S6.2.  KISSR galaxies indicated by red circles are luminous blue compact galaxies (LBCGs) identified by \citet{werk2004}.  Note that most of the LBCGs are located along the bottom of the left-hand figure at low [O III] EW.  Blue circles denote the low-redshift KISS galaxies that fall within the parameter space envelope defined by the \citet{Cardamone2009} GPs in all three comparison plots discussed in \S3.  Objects with double blue circles are low-redshift (z $<$ 0.09) GP candidates.}
\label{fig:lbcg_kiss}
\end{figure*}

In order to put this derived number density into context, we need to compare it to a comprehensive sample of galaxies located at a similar distance.  The luminosity function (LF) analysis presented in \citet{faber2007} for the DEEP2 redshift survey provides an excellent comparison sample of galaxies.   \citet{faber2007} derived robust LFs for both the DEEP2 and Combo-17 data sets in a series of redshift intervals.  We use their results for DEEP2 in the redshift range 0.2 -- 0.4, which is a reasonable match for the redshift range of the KISSR GPs of 0.29 -- 0.42.  Integrating their derived LF for all galaxies over the M$_B$ range $-$20.2 to $-$22.2 (i.e., the luminosity range of the KISSR GPs), we obtain a number density for the full galaxy population in this redshift range of 1.71 $\times$ 10$^{-3}$ Mpc$^{-3}$.  This is a factor of more than 1200 times larger that the number density of the KISSR GPs!  Clearly these GP-like galaxies are extremely rare, even at z = 0.3 -- 0.4.

We utilize our derived number density for the KISSR GP-like galaxies in the following section, where we attempt to identify nearby examples of the GP galaxies.

\section{Are There Local Analogs to the Green Peas?}
\label{sec:GPsNow}

Our discussion of the nature of the GPs presented in \S4 paints the picture of an extreme population of galaxies which presumably evolves extremely rapidly from z = 0.2 -- 0.4 to today.  It raises a number of key questions.   Do we see examples of the GPs in the very local Universe (z $<$ 0.1)?  What do the extreme GPs seen in KISSR look like today?  In this section we explore these questions by utilizing both of the low-redshift KISS samples of strong emission-line galaxies: the H$\alpha$-detected KISSR galaxies and the [O III]-selected KISS blue (hereafter KISSB) galaxies.

The question of whether we see GP galaxies in the local Universe is at least partially answered by the \citet{Cardamone2009} study, which detects GPs at redshifts as low as 0.141 (see \S3).  Since their color-selection method imposes strict redshift limits on their sample, there is no reason to expect that similar objects would not be present at even lower redshifts.  
We note that the shape the \citet{Cardamone2009} redshift histogram (Figure~\ref{fig:zHist}) {\it may} imply that GPs are less common at lower redshifts (z $<$ 0.225) than they are at higher redshift.  The survey volume difference between the redshift bin with the highest number of detections (0.275 to 0.30) and the bin at 0.15 exactly accounts for the difference in the number of detections.  However, the nearer redshifts should be sensitive to lower luminosity GPs, such that the expectations for detected sources in these bins would be higher than the predictions based purely on the geometry of the survey volume.  Taken on face value, this seems to imply a fading of the GP population.  Unfortunately, a more quantitative analysis would require a more precise definition of the selection function for the color-selected GP than is currently available.

On the other hand, Figures~\ref{fig:EWvSFR}-\ref{fig:AbvSFR} reveal that there are NO H$\alpha$-detected KISSR galaxies that overlap the properties of the [O~III]-detected KISSR galaxies.  While there are a handful of the lower-redshift KISSR galaxies that fall within the periphery of the \citet{Cardamone2009} GPs in these three figures (see below), none have properties that overlap with the higher-redshift KISSR galaxies.  This does not necessarily imply that such extreme objects do not exist in the local Universe, however.  The volume density of the [O~III]-detected KISSR GP-like galaxies (\S 5) combined with the effective volumes of the low-redshift portion of the KR1 and KR2 surveys ($\sim$520,000 Mpc$^3$) predicts that there should be 0.70 such galaxies within the H$\alpha$-detected portion of KISSR.   Hence, the rarity of the extreme GPs represented by the KISSR [O~III]-detected sample does not require that even a single such object be located within the H$\alpha$-detected KISSR catalog.  However, wide-field redshift surveys like 2dF \citep{colless2001} and SDSS \citep{sdssspec} should have detected dozens of such objects with z $<$ 0.10 if they exist in the local Universe.   We are not aware of any such detections.

In the following sub-sections we utilize the full H$\alpha$-detected KISSR catalog to look for possible GP-like galaxies in the local Universe.

\begin{figure*}[ht]
\centering
\includegraphics[width=\linewidth]{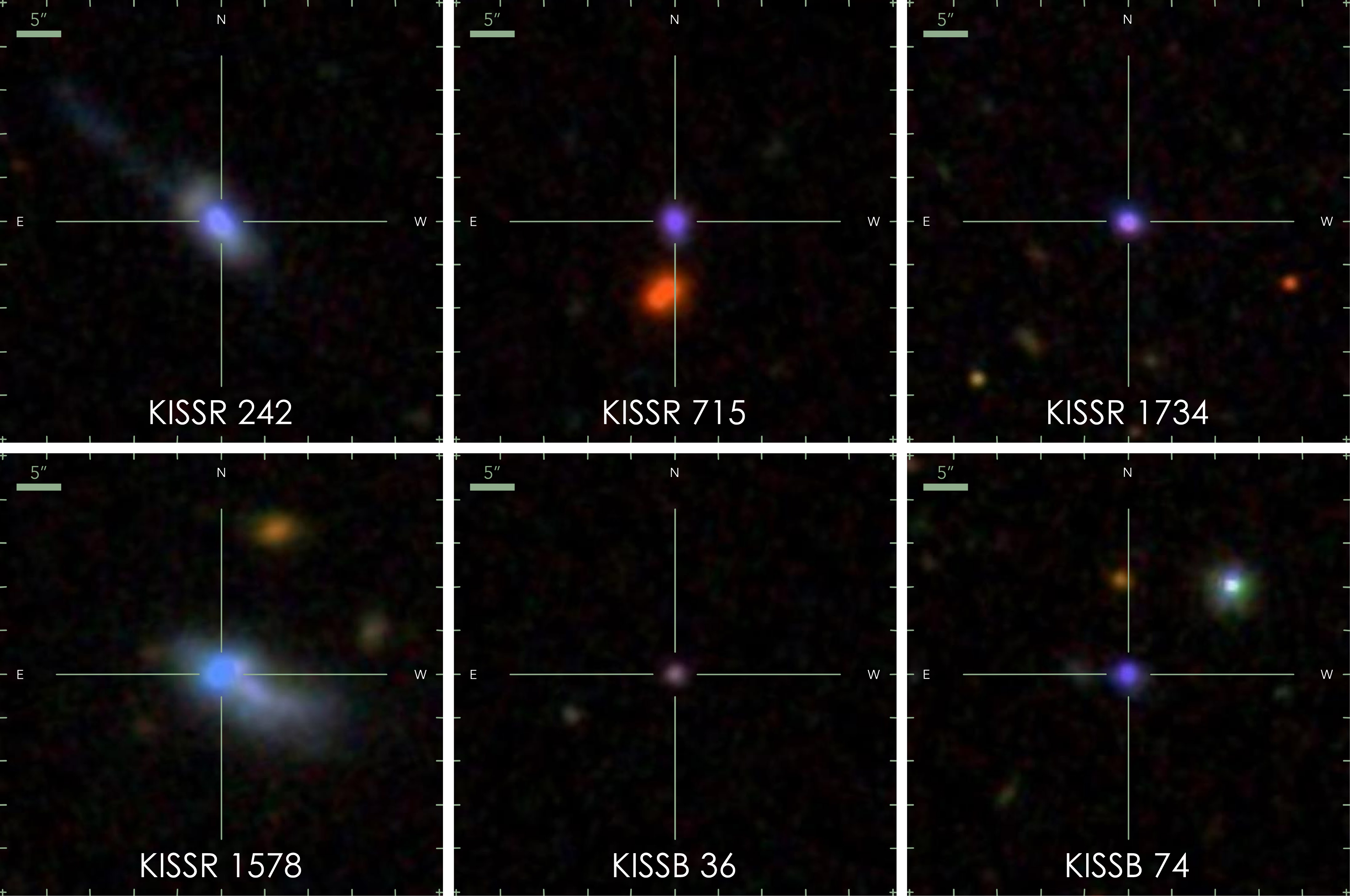}
\caption{\footnotesize \textit{Left:} SDSS images of two LBCGs (KISSR 242 and 1578) from the \citet{werk2004} sample.  Both LBCGs have a bright knot of star-formation embedded in a more extended galaxy and display disturbed morphologies.  Note the faint tidal feature extending above and to the left from KISSR 242. \textit{Middle and Right:} SDSS images of four galaxies from the KISS red (KISSR 715 and 1734) and KISS blue (KISSB 36 and 74) star-forming galaxy samples.  These galaxies are found in the parameter space occupied by the \citet{Cardamone2009} Green Peas, and they have similar compact morphologies as the \citet{Cardamone2009} and KISSR Green Peas.  The specific redshifts of these galaxies coupled with their strong emission lines result in their purple color (see text). 
}
\label{fig:analogs}
\end{figure*}

\subsection{Comparison with the Werk et al. Luminous Blue Compact Galaxies}

A natural starting point for our search for the Green Peas among the low-redshift KISS galaxies is to consider the properties of the KISS galaxies that have already been identified as being compact: the \citet{werk2004} sample of Luminous Blue Compact galaxies (LBCGs). 
For her study, \citet{werk2004} used the same selection criteria proposed by \citet{guzman2003}: Surface brightness (B-band) $<$ 21 mag~arcsec$^{-2}$, $M_{B}$ $<$ -18.5, and $B-V$ $<$ 0.6. 
It is important to note that the \citet{werk2004} LBCG sample is also redshift limited (z $<$ 0.045).  This constraint was imposed due to the low resolution of the KISS survey images.  In order to avoid including active galactic nuclei in the sample, \citet{werk2004} also required that the each galaxy selected as a LBCG had a follow-up spectrum that confirmed it as being star forming. This is an additional limitation because the spectral followup for the KISS survey was far from complete in 2004.  

To compare the properties of \citet{werk2004} LBCGs and our KISSR Green Peas, we followed the same procedure as we did in Section~\ref{sec:KISSasGP} when we compared the KISSR Green Peas to the \citet{Cardamone2009} Green Pea sample.  That is, we are looking to see if the LBCGs overlap the combined parameter space of the KISSR and \citet{Cardamone2009} Green Peas.  Figure~\ref{fig:lbcg_kiss} shows reproductions of Figures~\ref{fig:EWvSFR} and \ref{fig:MvsSFR} with the \citet{werk2004} LBCGs shown as black points with red circles around them.  

 Few of the LBCGs shown in Figure~\ref{fig:lbcg_kiss} reside in the same portion of parameter space as the Green Peas. There is nothing about the majority of the LBCGs that distinguishes them from the rest of the KISS H$\alpha$-selected star-forming galaxies; they have lower [O~III] EW, SFR, and sSFR compared with the Green Peas.  Most have EW$_{[O~III]}$ less than 50 \AA, which places them along the bottom of the left-hand plot in Figure~\ref{fig:lbcg_kiss}, while in the M$_*$--sSFR plot they mix well with the rest of the H$\alpha$-detected KISSR galaxies.
 
\begin{deluxetable*}{cccccccccccc}
\tabletypesize{\footnotesize}
\tablewidth{0pt}
\tablecaption{Properties of Low-Redshift KISS Green Pea Candidates \label{tab:lzprop}}

\tablehead{
 \colhead{KISSR} & \colhead{KISSB} & \colhead{B$_o$} &  \colhead{z} & \colhead{EW$_{[O~III]}$} & \colhead{M$_B$} & \colhead{log(M$_*$)} & \colhead{log(O/H)+12} & \colhead{log(SFR)} & \colhead{SF Age} & \colhead{SBE\tablenotemark{a}} & \colhead{\% SB Light}  \\
  &  &  &  & \AA &  &  \colhead{M$_\odot$}  &  & \colhead{M$_\odot$/yr} & \colhead{Myr} & mag \\
 (1)  & (2)  & (3)  & (4)  & (5)  & (6)  & (7)  & (8)  & (9)  & (10) & (11) & (12)
}
 
\startdata
  --   & \ 36 & 20.40 & 0.08874 & \ 888.7 & -17.64 & 8.55 & 7.68 & 0.40 & 141 & 0.84 & 55.9 \\
  --   & \ 74 & 18.86 & 0.05575 & \ 404.1 & -18.12 & 8.26 & 8.01 & 0.65 & \ 40  & 1.80 & 80.9\\
 \ 715 &  190 & 18.67 & 0.06815 &  1092.0 & -18.79 & 8.64 & 7.72 & 0.42 & 169  & 1.83 & 81.5\\
  1734 &  --  & 18.40 & 0.06568 & \ 548.7 & -18.97 & 9.64 & 8.26 & 0.72 & 845  & 0.34 & 26.9\\ 
\enddata
\tablenotetext{a}{SBE = Starburst Enhancement; see \S4}
\end{deluxetable*}
 
 There are two \citet{werk2004} LBCGs that do lie within the Green Pea galaxy parameter space in Figure~\ref{fig:lbcg_kiss}.  These two galaxies, KISSR 242 and 1578, while located at the less extreme end of the Green Peas with regard to their properties, still appear nonetheless to qualify as GPs.  However, when we look at their morphologies, we immediately are forced into the opposite conclusion.  Green Peas have a distinct semi-stellar morphology.  Example images of some of the KISSR Green Peas are shown in Figure~\ref{fig:images}.   The two LBCGs are shown in the left-most images in Figure~\ref{fig:analogs}, and the differences in morphology are clear.  The LBCGs are not compact in the same way as the Green Peas.  The two LBCGs have a compact, high surface brightness region of star formation embedded in a larger more extended galaxy.  These two systems are most likely mergers.  They are compact relative to the rest of the KISS H$\alpha$-selected star-forming galaxies, but they are much less compact than the KISSR Green Peas.  We note that this apparent difference in morphology could be due in part to redshift and/or resolution differences of the samples; it may well be that the extended structures and tidal features of these LBCGs would be undetectable in ground-based images if these galaxies were viewed at redshifts of 0.3-0.4.  However, we note that the other GP-like galaxies illustrated in Figure~\ref{fig:lbcg_kiss} (see \S 6.2) are located at distances of only a factor of $\sim$2 larger, yet show no extended structures.  The KISSR LBCGs are dramatically less compact than, for example, KISSR 1734, which has a similar stellar mass and SFR.  Hence, we conclude that the list of \citet{werk2004} LBCGs does not harbor any GP candidates.

\subsection{Searching for GP Candidates in the full KISS sample}

The evaluation of the KISS LBCGs did not result in the identification of any low-redshift GP-like objects.  However, the approach allowed us to recognize that some star-forming galaxies exhibit many properties similar to the GP galaxies without being compact enough to allow them to be classified as such.  Armed with this insight, we examined the properties of additional low-redshift KISSR and KISSB galaxies that happen to be located in the portion of parameter space occupied by the \citet{Cardamone2009} GP galaxies.

Our approach for trying to identify additional GP-like galaxies in the low-redshift portion of KISS was both simple and direct.  We identified those KISS galaxies located in the region of overlap with the \citet{Cardamone2009} GPs in Figures~\ref{fig:EWvSFR}-\ref{fig:AbvSFR} {\it and} which have compact morphologies.  
The objects with blue circles in Figure~\ref{fig:lbcg_kiss}, plus the LBCGs KISSR 242 and 1578 discussed in the previous subsection, were the only KISS galaxies that overlapped the parameters of the GPs in all three plots.  As was already clear from our previous discussion, these low-redshift KISS galaxies are just barely overlapping the GPs; they are all found along the perimeter of the parameter space covered by the \citet{Cardamone2009} galaxies, having lower SFRs than most of the {\it bona fide} GPs.  

When we evaluated the morphologies of these low-redshift GP candidates, we found that they fell into two distinct categories.  Four of them were extremely compact and showed little or no extended structures outside of the bright central core.  These four are illustrated in Figure~\ref{fig:analogs}: KISSR 715 and 1734, and KISSB 36 and 74.  We consider these four objects to be low-redshift GP analogs.  All four exhibit a strong violet color in the SDSS three-color images earning them the nickname Purple Grapes.  While their SDSS colors are similar to KISSR 1038 (Figure~\ref{fig:images}), the reason for the purple color is somewhat different.  In the case of these low-z systems, the strong [O~III] lines are located with the SDSS g-band (mapped to blue) while the H$\alpha$ line is transitioning into the SDSS i-band (mapped to red), giving rise to the violet color.  At even lower redshifts, the H$\alpha$ line will fall entirely in the SDSS r-band with the [O~III] lines still being located in the g-band, resulting in a strong blue color \citep[a.k.a., blue berries;][]{yang2017b, hysu2018}.  This is the case for the starburst regions in the two lower-redshift KISS LBCGs in Figure~\ref{fig:analogs}.

The remaining low-redshift GP candidates are nearly all either clearly in interacting or merging systems, or exhibit morphological peculiarities indicative of a recent merger.  Apparently the star-formation levels and spectral signatures exhibited in at least some merging systems are similar to those found in GP galaxies.  We are not necessarily trying to draw a connection between the GPs and recent merger activity, since the GPs typically do not show evidence for such activity (e.g., no obvious tidal features).   Nonetheless, the merger of two gas rich systems represents a viable explanation for the extreme star formation activity found in GPs.

The four low-z GP candidates identified above are denoted in Figure~\ref{fig:lbcg_kiss} by double blue circles.   Their properties are listed in Table~\ref{tab:lzprop}, which presents much of the same information found in Tables~\ref{tab:prop} and \ref{tab:extreme} for the [O~III]-detected KISS galaxies.  Not surprisingly, the low-redshift GP candidates tend to be lower luminosity and lower mass galaxies compared to the [O~III]-detected KISSR GPs.  Three of the four have stellar masses lower than any of their higher-redshift counterparts.  They also have metal abundances that are low for their luminosities, although they are not as extreme as the [O~III]-detected KISS GPs in this regard.  Three have very low star-formation ages, with KISSB 74 being capable of forming its entire stellar popultion in only 40 Myr at its current SFR. Finally, these local GP candidates are not as extreme as the higher-redshift KISS GPs in terms of their starburst luminosity enhancements.  While KISSR 715 and KISSB 74 both exhibit luminosity enhancements of $\sim$1.8 magnitudes, KISSR 1734 has an enhancement of only 0.34 magnitude and a fraction of its luminosity coming from the current starburst of 27\%.

The four KISS galaxies highlighted in Table~\ref{tab:lzprop} appear to represent low-reshift examples of the GP galaxies.  They are all much less extreme than the [O~III]-detected KISS GPs that are the subject of this paper.  Whether they represent faded examples of the types of GPs seen at higher redshift or simply lower-mass examples (or a mix) remains to be determined.  What is clear is that GP-like systems are present in the local Universe at redshifts as low as $\sim$0.05.

\subsection{What do the extreme GPs look like today?}

The discussion of the previous section at least partially answers the question of whether there are GP-like galaxies in the very local Universe.  We next consider the question of what the galaxies in the KISSR [O~III]-detected GP sample might look like today.  They are quite extreme in terms of their properties at z $\sim$ 0.3-0.4.  Furthermore, they must evolve substantially in the intervening 3-4 Gyr, since there do not appear to be any systems with such extreme properties at z $<$ 0.1.  If there were, they would stand out in the wide-field redshift surveys like 2dF \citep{colless2001} and SDSS \citep{sdssspec}.

Here we consider two possible evolutionary pathways for the KISSR [O~III]-selected GPs.
One possibility is that the galaxies seen as GPs at z $\sim$ 0.3-0.4 are still compact, star-forming systems.  In other words, they have retained some of their gas and are still making stars.  However, we infer that their SF activity would need to have dropped to a lower level than in the past, as mentioned above.  Presumably they would need to fade substantially, which in turn would imply substantially lower SFRs at the current time.

One of the lower redshift KISSR galaxies may represent such an example of a faded GP.  KISSR 1734 (Figure~\ref{fig:analogs}, Table~\ref{tab:lzprop}) has a stellar mass comparable to those of the KISSR [O~III]-detected GPs, a very compact morphology, and a SFR that is substantially lower than those exhibited by the higher redshift GPs of similar mass (4 to 8 times lower than the four with similar masses).  While KISSR 1734 was just bright enough to be included in the SDSS spectroscopic survey, systems with somewhat lower current SFRs and/or somewhat higher redshifts would likely have been missed by the currently available wide-field redshift surveys.  Hence, one could imagine that a modest population of faded GPs, having morphologies similar to their higher redshift counterparts, could exist in the local Universe without attracting our attention.

The second possibility we consider is a scenario where the GPs seen at z $\sim$ 0.3-0.4 have been totally quenched, presumably by a combination of using up their gas via star formation and gas removal via supernovae explosions.  In this case, the evolving GP would fade rather quickly.  It is also likely that the system would assume a somewhat less compact configuration after the loss of its gas, although the degree to which the stellar component expands would probably not be too dramatic.   Hence, we would imagine a centrally concentrated stellar system with a dominant 3-4 year old stellar population.  

In this scenario the evolving GP galaxies may end up looking like compact E+A galaxies: stellar systems with an elliptical-like morphology with no current star formation but a strong spectral signature of a young stellar population \citep{zabludoff1996}.  Originally, E+A galaxies were thought to be formed in rich galaxy clusters, but they have also been found in the field and in small groups \citep[e.g.,][]{zabludoff1996,yang2008}.  A subsample of E+A galaxies with very compact morphologies has been recognized \citep{zahid2016}.  However, the  \citet{zahid2016} sample consists of galaxies with higher masses than the Green Peas.  Because most known GPs are unresolved, it's hard to compare the relative sizes of the \citet{zahid2016} compact E+A galaxies and GPs. It's possible that the post-burst GP evolution represents a (probably minor) pathway to the formation of intermediate-mass compact E+A galaxies. 

It is worth noting that the remnant GPs will likely not evolve substantially due to hierarchical merging.  We are seeing them as compact, possibly young, systems at z = 0.3-0.4 (and at lower redshifts in the \citet{Cardamone2009} sample).  This is well past the epoch where galaxy growth due to accretion was efficient.  Hence, it seems safe to assume that the GPs will not accrete enough material to evolve into, for example, disk galaxies like we see in the local Universe.  

The precise method by which the GPs evolve may well depend on their local environments.  Preliminary studies suggest that GPs tend to be located in low and intermediate density environments \citep{kurtz2016}.  Ongoing work by our group (Brunker et al., in preparation) is looking into the galaxian environments populated by GPs.  

\section{Conclusions}
\label{sec:conclusions}

The primary focus of this paper has been to explore the properties and evolutionary status of the  [O~III]-selected star-forming galaxies identified by \citet{Salzer2009} from the KISSR emission-line galaxy survey.  While the bulk of the KISSR galaxies were detected via their H$\alpha$ emission at low redshifts (z $<$ 0.095), the [O~III]-selected galaxies were detected in the redshift range 0.29 -- 0.42 when strong [O~III]$\lambda$5007 entered the KISSR spectral bandpass.  We present new ground-based NIR and Spitzer MIR and FIR fluxes for members of this sample, and derive essential properties such as stellar masses and star-formation rates.  

We then compared the properties of the KISSR [O~III]-detected galaxies with the color-selected Green Pea galaxy sample from \citet{Cardamone2009} in order to assess the degree to which the galaxies in the two samples are similar.  Parameters compared between the two samples include stellar mass, [O~III]$\lambda$5007 equivalent width, star-formation rate, specific star-formation rate, and metallicity.  We find that the properties of the KISSR [O~III]-detected sample fully overlap those of the \citet{Cardamone2009} Green Peas. We conclude that the KISSR [O~III]-detected galaxies are consistent with being drawn from the same population of galaxies as the \citet{Cardamone2009} Green Peas. 

After ascertaining that the [O~III]-detected KISSR star-forming galaxies are in fact Green Peas, we further explored their nature by comparing their properties to those of the KISSR H$\alpha$-detected star-forming galaxies.  In addition to the comparisons mentioned above, we compared the B-band luminosities  of the two samples as well as there relative locations in mass-metallicity,  luminosity-metallicity, and mass-luminosity  diagrams.   In all comparisons considered the KISSR Green Peas stand out as being extreme.

Similar to results from studies of other samples of Green Peas, the KISSR Green Peas exhibit extremely high SFRs and sSFRs for their masses.  Unlike some previous studies that have parameterized Green Peas as being dwarf galaxies, the KISSR Green Pea sample extends to intermediate masses above 10$^{10}$ M$_\odot$. The KISSR Green Peas are also metal poor when compared to others galaxies of comparable luminosity or mass.  They exhibit an extreme displacement from the KISSR H$\alpha$-detected galaxies in the luminosity-metallicity diagram.  However, our analysis suggests that the offset in metal abundance is only 0.39 dex when the luminosity enhancement is accounted for.  Our use of the mass-luminosity relation reveals that the [O~III]-detected KISSR galaxies exhibit extreme starburst enhancements averaging 3.08 magnitudes.  That is, the KISSR Green Peas are, on average, a factor of 17 times brighter than a typical star-forming galaxy of the same mass.  This luminosity enhancement equates to 94\% of the light coming from the starburst population.  

One of the most astounding sets of numbers, from our perspective, are the extremely low star-formation ages (i.e., inverse of the sSFR) for the KISSR Green Peas as a class.  With ages as low as 31 Myr and a median age of 158 Myr, this parameter suggests strongly that these Green Peas are undergoing their first major episode of star formation.  The existence of young galaxies in the nearby Universe has been discussed and debated for years \citep[e.g.,][]{ss72, it2004,mamon2020}.  In the case of the most extreme KISSR GP systems, it is hard to imagine that the observed starbursts are part of the ongoing evolution of an otherwise normal galaxy.  In any case, the short formation timescales for Green Peas implies that they would be natural sources of escaping ionizing radiation in the early Universe, since their evolutionary ages nicely coincide with the young age of the Universe during the epoch of re-ionization.

We calculated the number density of the KISSR Green Peas to be 1.35 $\times$ 10$^{-6}$ Mpc$^{-3}$, which shows the Green Peas are extremely rare. We predict there should be 0.70 Green Peas in the volume covered by the low-redshift KISSR H$\alpha$-detected survey.  We used the KISSR H$\alpha$-detected catalog to look for Green Pea analogs in the local Universe (z $<$ 0.1).  We found four Green Pea analogs that have properties which overlap with the \citet{Cardamone2009} Green Peas and similar compact morphologies as the Green Peas. These analogs are less extreme in their properties compared with the KISS Green Peas, but they show that Green Pea-like systems are present in the local Universe.   At least one of these may represent a faded version of the more extreme Green Peas seen at intermediate redshifts.

\acknowledgements
We acknowledge financial support from the National Science Foundation for the KISS project as well as for the subsequent follow-up spectroscopy campaign (NSF-AST-9553020, NSF-AST-0071114, and NSF-AST-0307766).  We are grateful to the Indiana University College of Arts and Sciences for their continued support of the WIYN Observatory.  We thank the staff of the WIYN Observatory for their excellent support during both our WHIRC NIR imaging observations and our Hydra spectroscopic runs.  This work is based in part on observations made with the Spitzer Space Telescope, which was operated by the Jet Propulsion Laboratory, California Institute of Technology under a contract with NASA. This publication makes use of data products from the Wide-field Infrared Survey Explorer (WISE), which is a joint project of the University of California, Los Angeles, and the Jet Propulsion Laboratory/California Institute of Technology, funded by the National Aeronautics and Space Administration.  Funding for the SDSS has been provided by the Alfred P. Sloan Foundation, the Participating Institutions, the National Science Foundation, the U.S. Department of Energy, the National Aeronautics and Space Administration, the Japanese Monbukagakusho, the Max Planck Society, and the Higher Education Funding Council for England. The SDSS Web Site is http://www.sdss.org/. The SDSS is managed by the Astrophysical Research Consortium for the Participating Institutions. The Participating Institutions are the American Museum of Natural History, Astrophysical Institute Potsdam, University of Basel, University of Cambridge, Case Western Reserve University, University of Chicago, Drexel University, Fermilab, the Institute for Advanced Study, the Japan Participation Group, Johns Hopkins University, the Joint Institute for Nuclear Astrophysics, the Kavli Institute for Particle Astrophysics and Cosmology, the Korean Scientist Group, the Chinese Academy of Sciences (LAMOST), Los Alamos National Laboratory, the Max-Planck-Institute for Astronomy (MPIA), the Max-Planck-Institute for Astrophysics (MPA), New Mexico State University, Ohio State University, University of Pittsburgh, University of Portsmouth, Princeton University, the United States Naval Observatory, and the University of Washington.  

\vspace{5mm}
\facilities{WIYN (Hydra and WHIRC), Spitzer, WISE}

\end{document}